\documentclass[12pt,leqno]{article}
\title{Phase transitions in self-gravitating systems. Self-gravitating 
fermions and hard spheres models}
\author{Pierre-Henri Chavanis}
\date{}

\usepackage{amsthm,amsmath,amssymb,a4wide,psfig}
\def\mb#1{\setbox0=\hbox{$#1$}\kern-.025em\copy0\kern-\wd0
\kern-0.05em\copy0\kern-\wd0\kern-.025em\raise.0233em\box0}

\begin{document}
\maketitle
\vspace*{-1cm}
\begin{center}
 Laboratoire de Physique Quantique,
Universit\'e Paul Sabatier,\\
118 route de Narbonne 31062 Toulouse, France.\\

\vspace{0.5cm}
\end{center}

\begin{abstract}

We discuss the nature of phase transitions in self-gravitating systems
both in the microcanonical and in the canonical ensemble. We avoid the
divergence of the gravitational potential at short distances by
considering the case of self-gravitating fermions and hard spheres
models. Depending on the values of the parameters, three kinds of
phase transitions (of zeroth, first and second order) are
evidenced. They separate a ``gaseous'' phase with a smoothly varying
distribution of matter from a ``condensed'' phase with a core-halo
structure. We propose a simple analytical model to describe these
phase transitions. We determine the value of energy (in the
microcanonical ensemble) and temperature (in the canonical ensemble)
at the transition point and we study their dependance with the
degeneracy parameter (for fermions) or with the size of the particles
(for a hard spheres gas). Scaling laws are obtained analytically in
the asymptotic limit of a small short distance cut-off. Our analytical
model captures the essential physics of the problem and compares
remarkably well with the full numerical solutions. We also stress some
analogies with the liquid/gas transition and with the
Blume-Emery-Griffiths (BEG) model with infinite range interactions. In
particular, our system presents two tricritical points at which the
transition passes from first to second order.

\end{abstract}

\section{Introduction}
\label{sec_introduction}

The statistical mechanics of self-gravitating systems turns out to be
very different from that of other, more familiar, many-body systems
like neutral gases and plasmas due to the unshielded, long-range
nature of the gravitational force \cite{pad}. Because of this
fundamental difference, the notion of equilibrium is not always
well-defined and these systems exhibit a non trivial behavior with the
occurence of phase transitions associated with gravitational
collapse. If the particles are treated as classical point-masses, it
can be shown that no global entropy maximum exists, even if the system
is restricted within a box so as to prevent evaporation
\cite{antonov,lbw}. A self-gravitating system can increase entropy
without bound by developing a dense and hot ``core'' surrounded by a
dilute ``halo''. There exists, however, local entropy maxima
(metastable equilibrium states) if the condition $\Lambda=-{ER\over
GM^{2}}\le 0.335$ is satisfied, i.e. if the energy $E$ is sufficiently
large (for a given box radius $R$) or if the radius is sufficiently
small (for a given energy $E$). Since these equilibrium states are
only {\it local} entropy maxima, the question naturally emerges
whether they are long-lived or if they will collapse to a
configuration with higher entropy. In any case, a phase transition
{\it must} occur for $\Lambda>\Lambda_{c}=0.335$ since the entropy has
no extremum at all above this threshold \cite{antonov}. In that case,
the system is expected to collapse indefinitly towards a state of
higher and higher central concentration and temperature. This is the
celebrated ``gravothermal catastrophe'' \cite{lbw}.

However, if we introduce a repulsive potential at short distances,
complete core collapse is prevented and it can be proved that a global
entropy maximum now exists for all accessible values of energy. This
effective repulsion can be introduced in many different ways but the
physical results are rather insensitive to the precise form of the
regularization. For example, we can study the case of self-gravitating
fermions for which an exclusion principle imposes an upper bound on
the distribution function
\cite{hertel,bilic,cs,robert}. Alternatively, we can consider a
classical hard spheres gas by introducing an ``excluded volume''
around each particle \cite{aronson,stahl}. Other forms of
regularization are possible \cite{laliena,miller}.  When such
regularizations are introduced, it is possible to evidence properly
three kinds of phase transitions of zeroth, first and second
order. They separate a ``gaseous'' phase, which is independant on the
small scale cut-off, from a ``condensed'' phase in which the particles
are tightly bound. This is similar to the liquid/gas transition in an
ordinary fluid. However, for long-range systems such as
self-gravitating systems, the statistical ensembles are not
interchangeable and phase transitions can occur both in the canonical
and in the microcanonical ensemble. This results in the existence of
{\it two} tricritical points, one in each ensemble. In that respect the
self-gravitating Fermi gas shares some analogies with the
Blume-Emery-Griffiths (BEG) model with infinite range interactions
\cite{barre}. 

The object of this paper is to provide a detailed description of phase
transitions in self-gravitating systems. In the first part of the
paper, we consider the case of self-gravitating fermions. The
equilibrium phase diagram was calculated in an earlier paper \cite{cs}
and we complete this study by determining explicitly the values of
energy (in the microcanonical ensemble) and temperature (in the
canonical ensemble) at which the phase transitions occur. We also
propose a simple analytical model to describe these phase
transitions. The ``gaseous'' phase is modeled by a classical
homogeneous sphere while the ``condensed'' phase is made of a
completely degenerate nucleus surrounded by a hot atmosphere with
uniform density (restrained by the box). The mass $M_{*}$ of the nucleus is
determined by maximizing the entropy (resp. free energy) versus $M_{*}$
for a given total mass and energy (resp. temperature) of the configuration.
Quite remarkably, this simple model can reproduce the main features of
the numerical study. It also allows us to determine analytically how
the energy or the temperature at the transition points depend on the
degeneracy parameter. In the second part of the paper, we extend our
analytical model to the case of a classical gas with a short distance
cut-off. This model has been studied numerically by Aronson \& Hansen
\cite{aronson} and Stahl {\it et al.} \cite{stahl}, and our analytical
model gives a good agreement with their numerical results. It is also
consistent with the toy models of Lynden-Bell \& Lynden-Bell \cite{ll}
and Padmanabhan \cite{pad}. The possible astrophysical applications of
our study are discussed in Sec. \ref{sec_astro}.

\section{Statistical mechanics of self-gravitating fermions}
\label{sec_fermions}

\subsection{The Fermi-Dirac distribution}
\label{sec_fd}

We consider a system of $N$ fermions interacting via Newtonian
gravity. These particles can be electrons in white dwarf stars
\cite{chandra}, neutrons in neutron stars \cite{oppenheimer,hertel}, 
massive neutrinos in dark matter models
\cite{ruffini,bilic} etc... We assume that the mass of the configuration
is sufficiently small so as to ignore relativistic effects. Let $f({\bf r},{\bf
v},t)$ denote the distribution function of the system, i.e. $f({\bf
r},{\bf v},t)d^{3}{\bf r} d^{3}{\bf v}$ gives the mass of particles
whose position and velocity are in the cell $({\bf r},{\bf v};{\bf
r}+d^{3}{\bf r},{\bf v}+d^{3}{\bf v})$ at time $t$. The integral of
$f$ over the velocity determines the spatial density
\begin{equation}
\rho=\int f d^{3}{\bf v},
\label{fd1}
\end{equation}
and the total mass of the configuration is given by
\begin{equation}
M=\int \rho d^{3}{\bf r},
\label{fd2}
\end{equation}
where the integral extends over the entire domain. On the other hand,
in the meanfield approximation, the total energy of the system can be
expressed as
\begin{equation}
E={1\over 2}\int fv^{2}d^{3}{\bf r}d^{3}{\bf v}+{1\over 2}\int\rho\Phi d^{3}{\bf r}=K+W,
\label{fd3}
\end{equation}
where $K$ is the kinetic energy and $W$ the potential energy. The
gravitational potential $\Phi$ is related to the star density by the
Newton-Poisson equation
\begin{equation}
\Delta\Phi=4\pi G\rho.
\label{fd4}
\end{equation}
Finally, the Fermi-Dirac entropy is given by the formula
\begin{equation}
S=-\int \biggl\lbrace {f\over\eta_{0}}\ln {f\over\eta_{0}}+\biggl (1-  {f\over\eta_{0}}\biggr)\ln \biggl (1-  {f\over\eta_{0}}\biggr)\biggr\rbrace d^{3}{\bf r}d^{3}{\bf v}, 
\label{fd5}
\end{equation}
which can be obtained by a standard combinatorial analysis. In this
expression, $\eta_{0}$ is the maximum value accessible to the
distribution function. If $g=2s+1$ denotes the spin multiplicity of
the quantum states, $m$ the mass of the particles and $h$ the Planck
constant, one has by virtue of the Pauli exclusion principle
$\eta_{0}={gm^{4}/ h^{3}}$. An entropy of the form (\ref{fd5}) was
also introduced by Lynden-Bell \cite{lb} in the context of 
collisionless self-gravitating
systems (e.g., elliptical galaxies, dark matter) undergoing a
``violent relaxation'' by phase mixing
\cite{csr,kull,dubrovnik}. In that context, $\eta_{0}$ represents
the maximum value of the initial distribution function and the actual
distribution function (coarse-grained) must always satisfy
$\overline{f}\le \eta_{0}$ by virtue of the Liouville theorem. This is
the origin of the ``effective'' exclusion principle in Lynden-Bell's
theory which has obviously nothing to do with quantum mechanics. In
reality, the mixing entropy introduced by Lynden-Bell is a complicated sum of
Fermi-Dirac entropies for each phase level constituting the initial
condition. For simplicity, we shall restrict ourselves to the single
level approximation for which the mixing entropy coincides with
expression (\ref{fd5}).

At statistical equilibrium, the system is expected to maximize the Fermi-Dirac entropy at fixed mass and energy. Introducing Lagrange multipliers to satisfy these constraints, we find that the {\it critical points} of entropy correspond to the Fermi-Dirac distribution
\begin{equation}
f={\eta_{0}\over 1+\lambda e^{\beta ({v^{2}\over 2}+\Phi)}},
\label{fd6}
\end{equation}
where $\lambda$ is a strictly positive constant insuring that $f\le \eta_{0}$ and $\beta$ is the inverse temperature. In the fully degenerate limit $f\simeq \eta_{0}$, this distribution function has been extensively studied in the context of white dwarf stars in which gravity is balanced by the pressure of a degenerate electron gas \cite{chandra}. In the non degenerate limit $f\ll\eta_{0}$, the Fermi-Dirac distribution reduces to the Maxwell-Boltzmann distribution
\begin{equation}
f\simeq {\eta_{0}\over \lambda}e^{-\beta({v^{2}\over 2}+\Phi)},
\label{fd7}
\end{equation}
so that we expect to recover the properties of classical isothermal spheres at low densities \cite{pad}. In particular, the Fermi-Dirac spheres, like the isothermal spheres, have an infinite mass  and one is forced to confine the system within a box of radius $R$. Physically, this confinement is justified by the realization that the relaxation is {\it incomplete} so that the conditions of applicability of statistical mechanics, regarding for example the ergodic hypothesis, can be fulfilled only in a limited region of space. In addition, an astrophysical system is never completely isolated and $R$ could represent the typical radius at which the system interacts with its neighbors. 

Kinetic equations have been proposed to describe the relaxation of self-gravitating systems towards the Fermi-Dirac distribution \cite{csr,quasi}. If we assume that the system is subject to tidal forces and if we allow high energy particles to escape the system when they reach an energy $\epsilon={v^{2}\over 2}+\Phi\ge \epsilon_{m}$, an extension of the Michie-King model taking into account the degeneracy can be deduced from these equations \cite{quasi}. For $\epsilon\le \epsilon_{m}$, one has   
\begin{equation}
f=\eta_{0}{e^{-\beta\epsilon}-e^{-\beta\epsilon_{m}}\over \lambda+e^{-\beta\epsilon}},
\label{fd8}
\end{equation} 
while $f=0$ for $\epsilon>\epsilon_{m}$ since the stars have been removed by the tidal field. When $\lambda\rightarrow +\infty$, we recover the Michie-King model \cite{bt} and when $\epsilon_{m}\rightarrow +\infty$, we recover the Fermi-Dirac distribution (\ref{fd6}). The density associated with this distribution function goes to zero at a finite radius, identified as the tidal radius. Therefore, the configuration has a finite mass. This distribution function could describe elliptical galaxies and galactic halos limited in extension as a consequence of tidal interactions with other systems \cite{lb,ruffini}. This model is of course more realistic than the box model. However, in order to exhibit phase transitions in self-gravitating systems, the box model provides a more convenient theoretical framework and we shall use it in the sequel.

\subsection{Thermodynamical parameters}
\label{sec_para}

The thermodynamical parameters for Fermi-Dirac spheres in the
meanfield approximation have been calculated by Chavanis \& Sommeria
\cite{cs} and we shall directly use their results. The 
equation determining the gravitational potential at equilibrium is
obtained by substituting the Fermi-Dirac distribution (\ref{fd6}) in
the Poisson equation (\ref{fd4}), using Eq. (\ref{fd1}).  Introducing
the variables $\psi=\beta(\Phi-\Phi_{0})$, where $\Phi_{0}$ is the
central potential, $k=\lambda e^{\beta\Phi_{0}}$ and
$\xi=({16\pi^{2}\sqrt{2}G\eta_{0}/ \beta^{1/2}})^{1/2}r$, it can be
written
\begin{equation}
{1\over\xi^{2}}{d\over d\xi}\biggl (\xi^{2}{d\psi\over d\xi}\biggr )=I_{1/2}(ke^{\psi(\xi)}),
\label{p1}
\end{equation}
where $I_{1/2}$ denotes the Fermi integral
\begin{equation}
I_{n}(t)=\int_{0}^{+\infty}{x^{n}\over 1+t e^{x}}dx,
\label{p2}
\end{equation} 
of order $n=1/2$. The boundary conditions at the origin are
\begin{equation}
\psi(0)=\psi'(0)=0.
\label{p3}
\end{equation} 
In the case of bounded spheres, one must stop the integration of Eq. (\ref{p1}) at $\xi=\alpha$ with
\begin{equation}
\alpha=\biggl ({16\pi^{2}\sqrt{2}G\eta_{0}\over \beta^{1/2}}\biggr )^{1/2}R.
\label{p4}
\end{equation} 
The parameter $\alpha$ is related to the temperature and
to the energy by the equations
\begin{equation}
\eta\equiv {\beta GM\over R}=\alpha\psi'_{k}(\alpha),
\label{p5}
\end{equation} 
\begin{equation}
\Lambda\equiv -{ER\over GM^{2}}={\alpha^{7}\over \mu^{4}}\int_{0}^{\alpha}I_{3/2}(ke^{\psi_{k}(\xi)})\xi^{2}d\xi-{2\over 3}{\alpha^{10}\over \mu^{4}}I_{3/2}(ke^{\psi_{k}(\alpha)}).
\label{p6}
\end{equation} 
Moreover, $\alpha$ and $k$ are related to each other by the relation 
\begin{equation}
\alpha^{5}\psi'_{k}(\alpha)=\mu^{2},
\label{p7}
\end{equation} 
where $\mu$ is the ``degeneracy parameter''
\begin{equation}
\mu=\eta_{0}\sqrt{512\pi^{4}G^{3}MR^{3}}.
\label{p8}
\end{equation} 
For a given value of $\mu$ and $k$, we can solve the ordinary differential equation (\ref{p1}) until the value $\xi=\alpha$ at which the condition (\ref{p7}) is satisfied. Then, Eqs. (\ref{p5})(\ref{p6}) determine the temperature and the energy of the configuration. By varying the parameter $k$ (for a fixed value of the degeneracy parameter $\mu$), we can cover the whole bifurcation diagram in parameter space \cite{cs}. The entropy of each configuration is given by (see Appendix A)
\begin{equation}
{S\eta_{0}\over M}=-{7\over 3}\Lambda\eta+\psi_{k}(\alpha)+\eta+\ln k-{2\alpha^{6}\over 9\mu^{2}}I_{3/2}(ke^{\psi_{k}(\alpha)}),
\label{p9}
\end{equation}
and the free energy by
\begin{equation}
J=S-\beta E.
\label{p10}
\end{equation} 
Note that Eq. (\ref{p10}) is the free energy $F=E-TS$ up to a negative
proportionality factor. In the microcanonical ensemble, a solution is
stable if it corresponds to a maximum of entropy $S$ at fixed mass and
energy. In the canonical ensemble, the condition of stability requires
that the solution be a maximum of free energy $J$ at fixed mass and
temperature. It can be shown \cite{hertel} that this meanfield
approach is {\it exact} in a thermodynamical limit such that
$N\rightarrow +\infty$ with $\mu$, $\eta$, $\Lambda$ fixed. This
implies in particular that $N^{1/3}R$, $TN^{-4/3}$,
$EN^{-7/3}$ and $SN^{-1}$ approach a constant value for $N\rightarrow
+\infty$. The usual thermodynamical limit $N,R\rightarrow +\infty$ with
$N/R^{3}$ constant is of course not relevant for non extensive systems.

\section{Phase transitions in a self-gravitating Fermi gas}
\label{sec_pt}

\subsection{The non degenerate limit $(\mu=\infty)$}
\label{sec_class}

Before considering the case of an arbitrary degree of degeneracy, it
may be useful to discuss first the non degenerate limit corresponding
to a classical isothermal gas ($\hbar\rightarrow 0$). In that case, the
thermodynamical parameters are given by \cite{lbw}
\begin{equation}
\eta=\alpha\psi'(\alpha),
\label{cc1}
\end{equation} 
\begin{equation}
\Lambda={3\over 2}{1\over\alpha\psi'(\alpha)}-{e^{-\psi(\alpha)}\over\psi'(\alpha)^{2}},
\label{cc2}
\end{equation} 
\begin{equation}
{S\over N}=-{1\over 2}\ln\eta-2\ln\alpha+\psi(\alpha)+\eta-2\Lambda\eta,
\label{cc3}
\end{equation} 
where $\alpha=(4\pi G\beta\rho_{0})^{1/2}R$ is the normalized box radius and $\psi$ is the normalized gravitational potential solution of the Emden equation \cite{chandra}
\begin{equation}
{1\over \xi^{2}}{d\over d\xi}\biggl (\xi^{2}{d\psi\over d\xi}\biggr )=e^{-\psi},
\label{cc4}
\end{equation} 
with boundary conditions
\begin{equation}
\psi(0)=\psi'(0)=0.
\label{cc5}
\end{equation}

\begin{figure}[htbp]
\centerline{
\psfig{figure=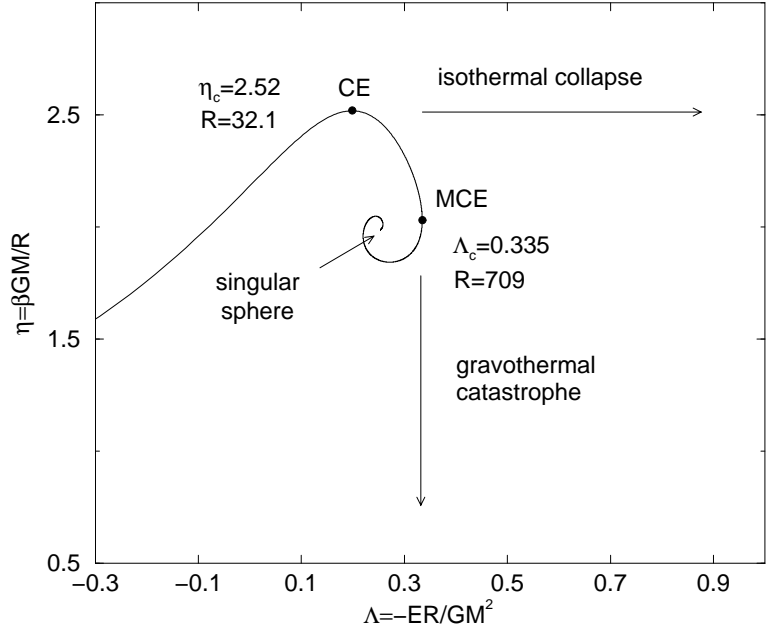,angle=0,height=8.5cm}}
\caption{Equilibrium phase diagram for classical isothermal spheres. For $\Lambda>\Lambda_{c}$ or $\eta>\eta_{c}$, there is no hydrostatic equilibrium and the system undergoes a gravitational collapse.}
\label{etalambda}
\end{figure}

\begin{figure}[htbp]
\centerline{
\psfig{figure=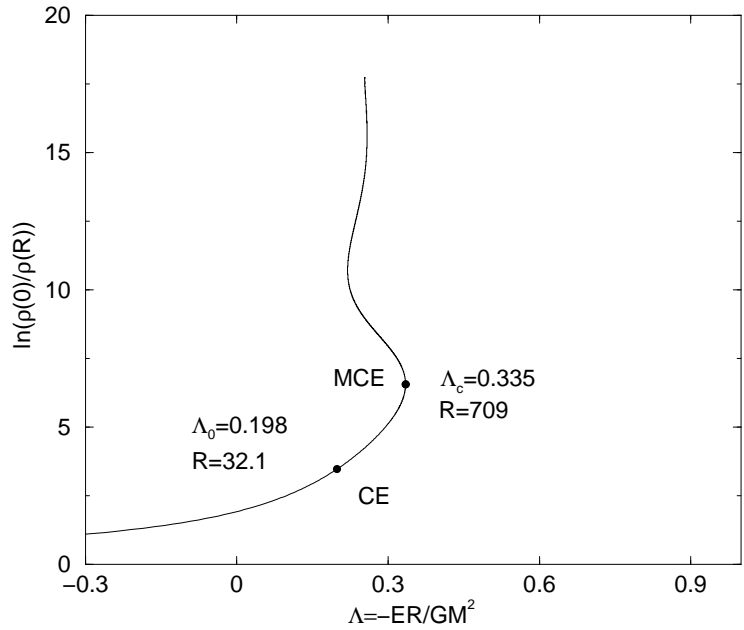,angle=0,height=8.5cm}}
\caption{Density contrast of classical isothermal spheres as a function of energy. The series of equilibrium becomes unstable in the canonical ensemble for ${\cal R}>32.1$ and in the canonical ensemble for ${\cal R}>709$. The value of energy at which the density contrast tends to $+\infty$ is $\Lambda=1/4$ (singular sphere).}
\label{contrast}
\end{figure}

The equilibrium phase diagram $(E,T)$ is represented in
Fig. \ref{etalambda}. The curve is parametrized by $\alpha$ which can
be considered as a measure of the central concentration. An equivalent
parametrization is provided by the density contrast ${\cal
R}=\rho(0)/\rho(R)$ (see Fig. \ref{contrast}). In the microcanonical
ensemble, the solutions on the upper branch of Fig. \ref{etalambda}
(until point MCE) are stable and correspond to local entropy
maxima. The rest of the spiral corresponds to unstable saddle
points. For $\Lambda>\Lambda_{c}=0.335$, there are no critical points
of entropy and the system collapses indefinitely. This is the
so-called ``gravothermal catastrophe''. In the canonical ensemble,
only the solutions prior to the point CE are stable. They correspond
to local maxima of free energy. The rest of the spiral corresponds to
unstable saddle points. For $\eta>\eta_{c}=2.52$, there is no
hydrostatic equilibrium and the system undergoes an ``isothermal
collapse''. These stability results can be deduced from the turning
point analysis of Katz \cite{katz} or by solving the eigenvalue
equations associated with the second order variations of entropy or
free energy \cite{pad2,chaviso}. For $\Lambda<\Lambda_{c}$ and
$\eta<\eta_{c}$, the stable solutions are only {\it metastable}. There
are no global maxima of entropy or free energy for classical point
masses in gravitational interaction \cite{antonov}.  We also note that
the region of negative specific heats (between points CE and MCE) is
stable in the microcanonical ensemble and unstable in the canonical
ensemble, where it is replaced by a phase transition (an ``isothermal
collapse''). This is expected on physical grounds since it can be
shown quite generally that the specific heat {\it must} be positive in
the canonical ensemble \cite{pad}. These results clearly indicate that
the statistical ensembles are not interchangeable in the case of
self-gravitating systems (see, e.g., Ref. \cite{crs}), contrary to
normal matter in which the energy is an extensive parameter.

\begin{figure}[htbp]
\centerline{
\psfig{figure=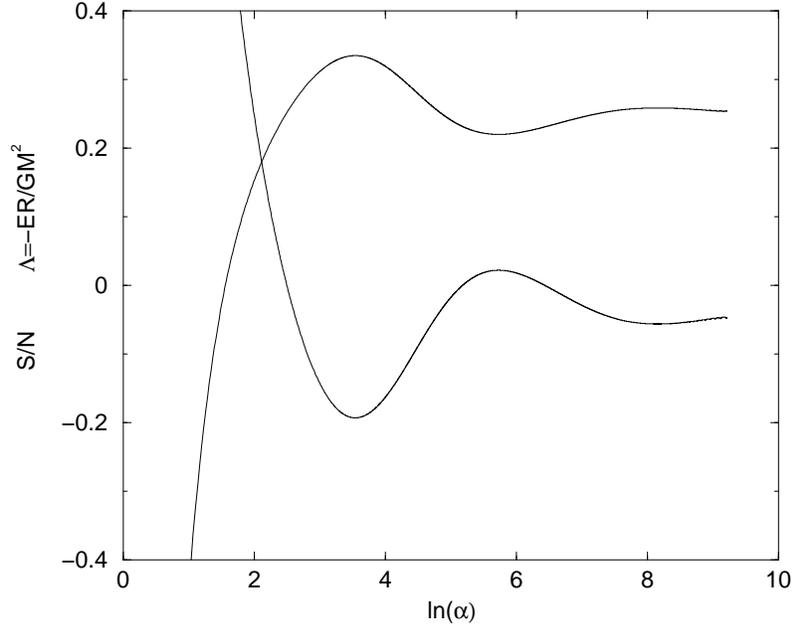,angle=0,height=8.5cm}}
\caption{Entropy and energy as a function of the central concentration $\alpha$. The peaks of energy and entropy occur for the same values of $\alpha$.}
\label{alphaSL}
\end{figure}

\begin{figure}[htbp]
\centerline{
\psfig{figure=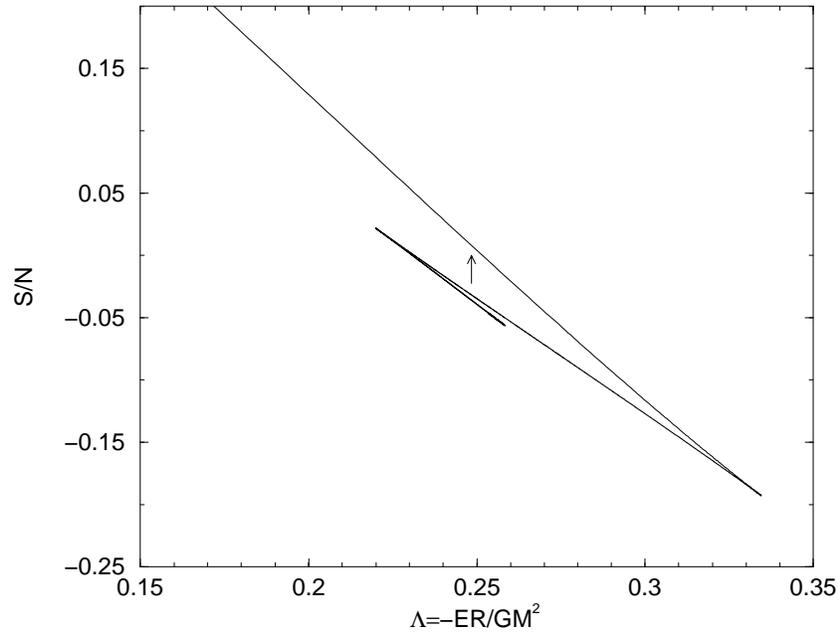,angle=0,height=8.5cm}}
\caption{Entropy versus energy for classical isothermal spheres. When several solutions exist for the same energy, the states with low entropy are unstable saddle points. They can either evolve towards the metastable state with highest entropy (see arrow) or collapse to a state of ever increasing entropy, as suggested in Ref. \cite{crs}.}
\label{LambdaS}
\end{figure}

\begin{figure}[htbp]
\centerline{
\psfig{figure=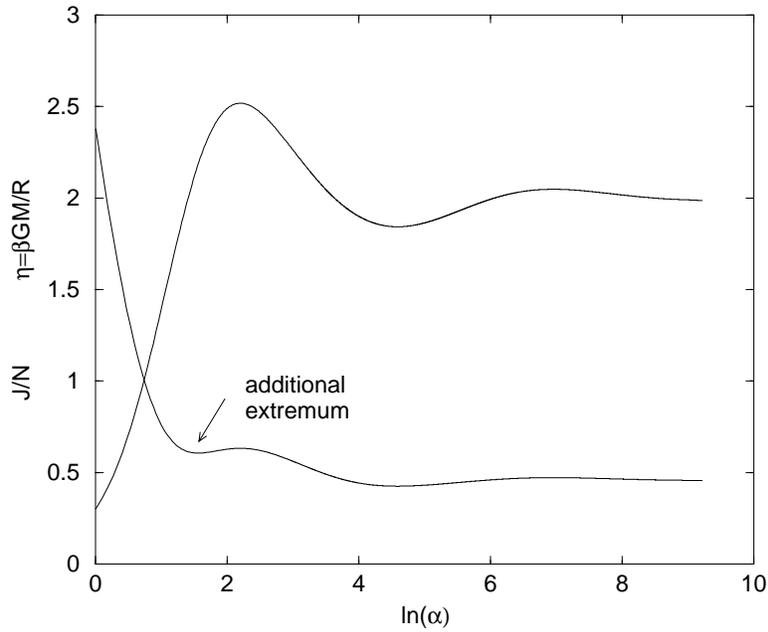,angle=0,height=8.5cm}}
\caption{Free energy and inverse temperature as a function of the central concentration $\alpha$. The peaks occur for the same values of $\alpha$. The free energy presents an additional extremum but it is not associated to an instability.}
\label{Jalpha}
\end{figure}

\begin{figure}[htbp]
\centerline{
\psfig{figure=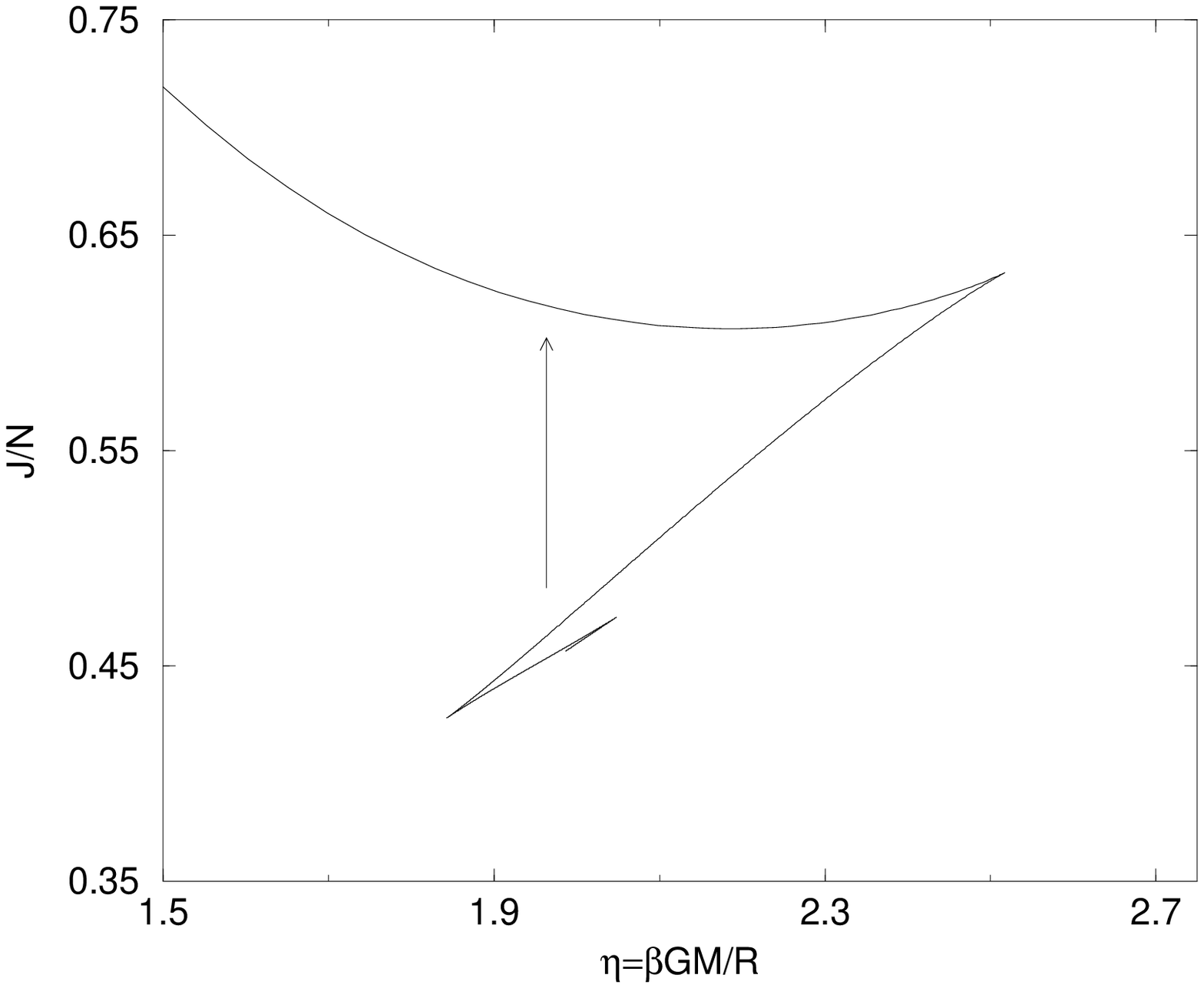,angle=0,height=8.5cm}}
\caption{Free energy versus inverse temperature for classical isothermal spheres. The interpretation is the same as in Fig. \ref{LambdaS}.}
\label{etaJ}
\end{figure}

In Fig. \ref{alphaSL}, we have ploted the energy and the entropy as a
function of the central concentration $\alpha$. We observe that the
peaks occur for the {\it same} values of $\alpha$. This is also clear
from Fig. \ref{LambdaS} where we have represented the entropy as a
function of the energy. When several critical points of entropy exist
for the same energy, only the one with the largest entropy is an
entropy {\it maximum}. The other critical points are unstable saddle
points. Therefore, if the system is initially prepared on a saddle
point, we expect a transition to occur from a state of low entropy to
a state of higher entropy. This is not really a phase
transition but just an instability. A similar diagram has been found
for isothermal spheres described in the context of general relativity
\cite{relat}. In this analogy, the mass-energy $M$ plays the role of
the classical energy $E$ and the binding energy
$E_{bind.}=(M-M_{0})c^{2}$, where $M_{0}=Nm$ is the rest mass and $N$
the baryon number, plays the role of the classical entropy $S$.  In
Figs. \ref{Jalpha}-\ref{etaJ}, we have ploted the corresponding
diagrams in the canonical ensemble. Here again, the peaks of
temperature correspond to the peaks of free energy. The free energy
has an additional peak which has no counterpart in the temperature
diagram. However, this peak is not associated to an instability and
the interpretation of the curves is the same as in the microcanonical
ensemble.

\subsection{Large values of the degeneracy parameter ($\mu=10^{5}$)}
\label{sec_large}

\begin{figure}[htbp]
\centerline{
\psfig{figure=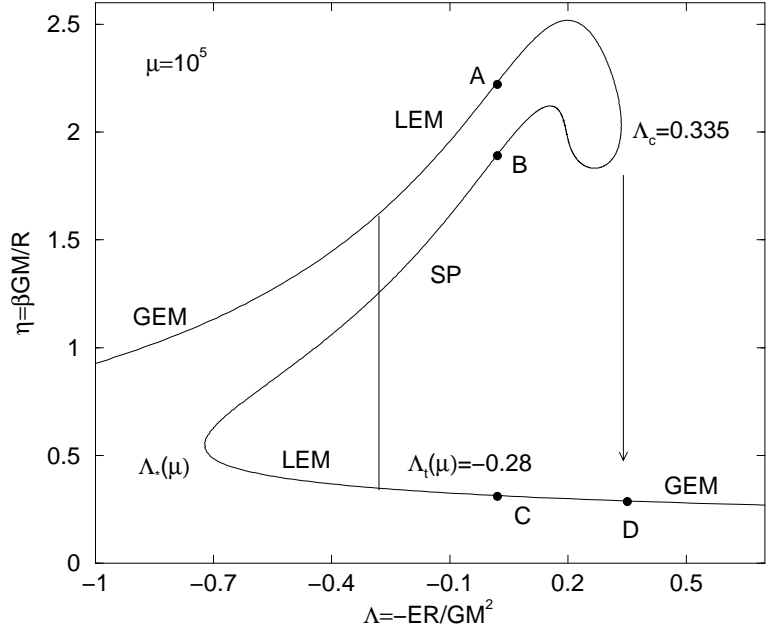,angle=0,height=8.5cm}}
\caption{Equilibrium phase diagram for Fermi-Dirac spheres with a degeneracy parameter $\mu=10^{5}$. Points A  form the ``gaseous''  phase. They are global entropy maxima (GEM) for $\Lambda<\Lambda_{t}(\mu)$ and local entropy maxima (LEM) for $\Lambda>\Lambda_{t}(\mu)$. Points C form the ``condensed'' phase. They are LEM for $\Lambda<\Lambda_{t}(\mu)$ and  GEM for $\Lambda>\Lambda_{t}(\mu)$. Points B are unstable saddle points (SP) and contain a ``germ''. This figure exhibits in particular a first order phase transition in the microcanonical ensemble.}
\label{le5}
\end{figure}

\begin{figure}[htbp]
\centerline{
\psfig{figure=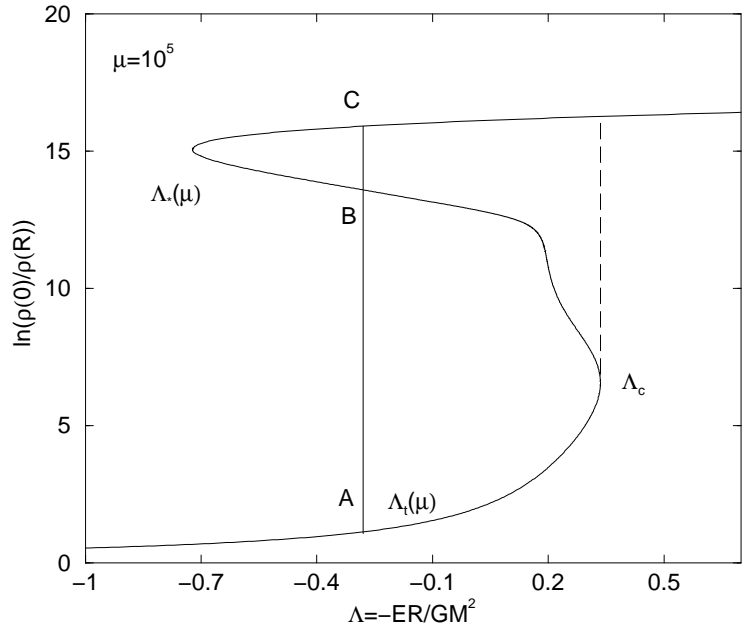,angle=0,height=8.5cm}}
\caption{Density contrast as a function of energy for self-gravitating 
fermions ($\mu=10^{5}$). This figure can be compared with Fig. \ref{contrast} for classical isothermal spheres. Points A (gaseous phase) have a low density contrast. Points B and C contain a central nucleus with high density \cite{cs}.  }
\label{contrast5}
\end{figure}

We now consider the case of self-gravitating fermions characterized by
a degeneracy parameter $\mu$. We first discuss the case of large
values of the degeneracy parameter. The extension of the classical
diagram of Fig. \ref{etalambda} is reported in Fig. \ref{le5}. We see
that the inclusion of degeneracy has the effect of unwinding the
spiral. The evolution of the density contrast along the series of
equilibrium is depicted in Fig. \ref{contrast5}. In the range
$\Lambda_{*}(\mu)<\Lambda<\Lambda_{c}$, there exists several critical
points of entropy for each single value of energy. The solutions on
the upper branch of Fig. \ref{le5} (points A) are non degenerate and
have a smooth density profile; they form the ``gaseous'' phase. The
solutions on the lower branch (points C) have a ``core-halo''
structure with a massive degenerate nucleus and a dilute atmosphere;
they form the ``condensed'' phase. According to the criterion of Katz
\cite{katz}, specifically discussed in Ref. \cite{cs} for
self-gravitating fermions, these solutions are both entropy maxima
(EM) while the intermediate solutions (points B) are unstable saddle
points (SP). These points are similar to points A, except that they
contain a small embryonic nucleus (with small mass and energy) which
plays the role of a ``germ'' in the langage of phase transition. The
density profiles of these solutions are given in Ref. \cite{cs}.

\begin{figure}[htbp]
\centerline{
\psfig{figure=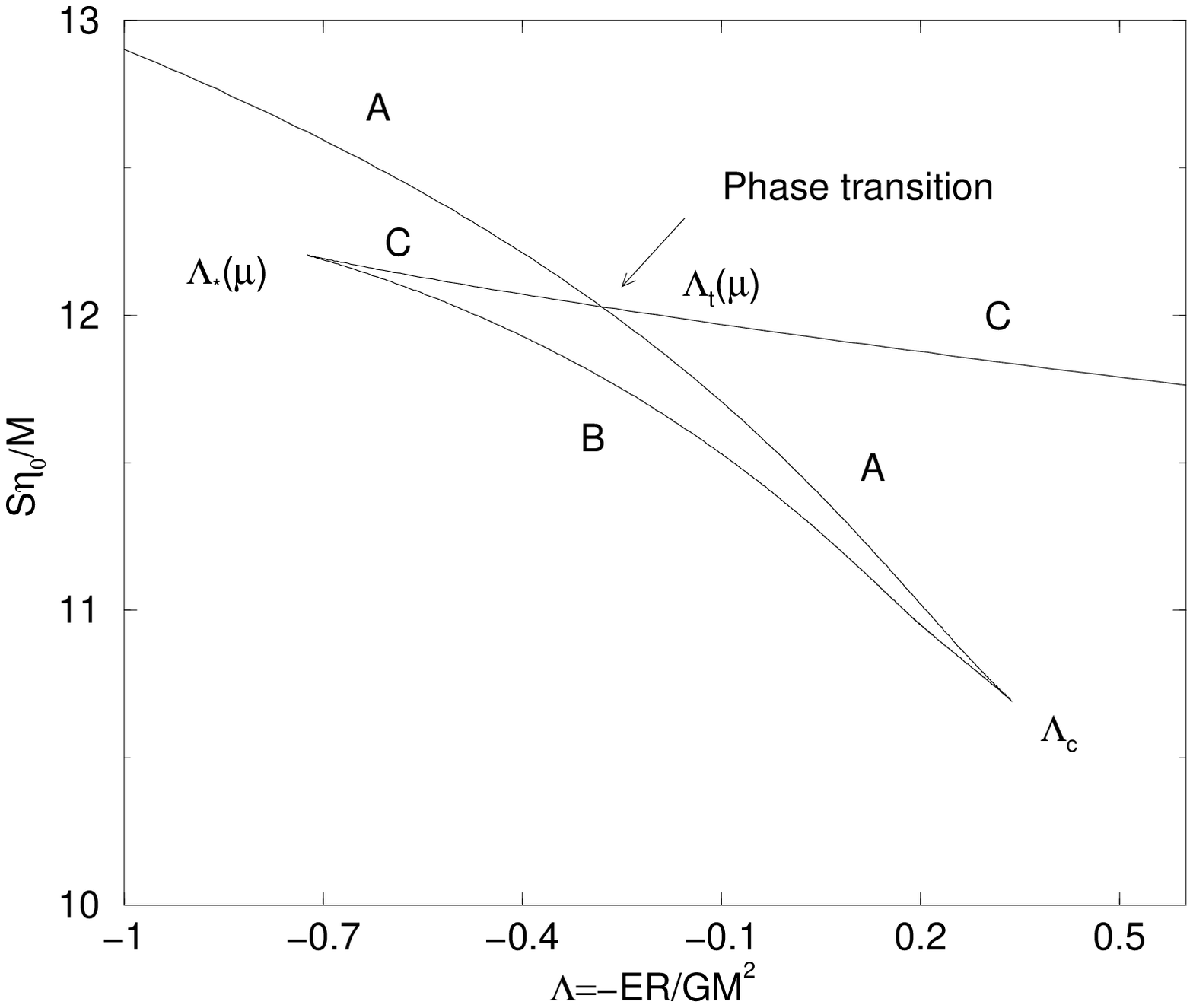,angle=0,height=8.5cm}}
\caption{Entropy of each phase versus energy for $\mu=10^{5}$. A phase transition occurs at $\Lambda_{t}(\mu)$ at which the two stable branches (solutions A and C) intersect. The unstable solutions B always have smaller entropy.}
\label{SL5}
\end{figure}

To be more precise, we have ploted the entropy of these solutions as a
function of energy in Fig. \ref{SL5}. The entropy of the unstable
phase (points B) is always smaller than the entropy of the stable
phases, as it should. There is now a crossing point in the diagram, at
$\Lambda=\Lambda_{t}(\mu)$, which marks the onset of a phase
transition. At that point, the ``gaseous'' phase and the ``condensed''
phase have the same entropy. As $\Lambda$ is increased across the
transition point, the non degenerate solutions (points A) pass from
global to local entropy maxima. Inversely, the degenerate solutions
(points C) pass from local to global entropy maxima.  We expect
therefore a phase transition to occur from the ``gaseous'' phase to
the ``condensed'' phase when $\Lambda=\Lambda_{t}^{+}(\mu)$. The
``kink'' in the curve $S(E)$ at the transition point where the two
branches intersect corresponds to a discontinuity of temperature in
the equilibrium phase diagram (see the vertical plateau in
Fig. \ref{le5}). The specific heat is also discontinuous at that point
and turns from positive to negative. According to Stahl {\it et al.} 
\cite{stahl}, this phase transition could be called a ``gravitational
first order phase transition''. It has to be noted, however, that
contrary to the liquid/vapor transition, the two phases cannot coexist
in the present situation.

For $\Lambda_{t}(\mu)<\Lambda<\Lambda_{c}$, the non degenerate
solutions are metastable but we may suspect \cite{cs} that they are
long-lived so that they {\it are} physical. These solutions are
insensitive to the small-scale regularization and depend only on the
long-range gravitational interaction. In the limit $\mu\rightarrow
+\infty$, the transition energy $\Lambda_{t}(\mu)$ goes to $-\infty$
and we recover the classical spiral of Fig. \ref{etalambda}. This
spiral is formed by the metastable states of the ``gaseous'' phase
(points A).  The ``condensed'' phase (points C) is superposed to the
$\Lambda$-axis. These states have an infinite central density and an
infinite temperature. The unstable branch (points B) coincides with
the spiral but these states physically differ from the ``gaseous''
states (points A) by the presence of an infinitesimal ``germ'' with
negligible mass and energy. Therefore, what we see actually in the
limit $\mu\rightarrow +\infty$ are the metastable states.  It is
plausible that these metastable states will be selected by the
dynamics (on relevant time scales) even if states with higher entropy
exist. In fact, depending on its {\it topology} (i.e., the form of the
profile) an initial condition with $\Lambda>\Lambda_{t}(\mu)$ can
either relax towards the local entropy maximum (gaseous phase) or
collapse towards the global entropy maximum (condensed phase).
Alternatively, for $\Lambda<\Lambda_{t}(\mu)$ an initially
``condensed'' configuration can remain frozen in this metastable state
or explose in a ``gaseous'' state with more
entropy. Therefore, the choice of a stable equilibrium state does not
only depend whether the equilibrium solution is a local or a global
entropy maximum. It is more dependant whether the initial condition
lies in the ``basin of attraction'' of the equilibrium state or
not. The characterization of this basin of attraction requires a
nonequilibrium analysis which is not attempted in the present paper. A
first step in that direction was performed by Youngkins \& Miller
\cite{miller} by using a one-dimensional spherical shell model and by
Chavanis {\it et al.} \cite{crs} with the aid of a simple relaxation
equation derived from a maximum entropy production principle
\cite{csr}. These preliminary works reveal that 
the structure of this basin of attraction is extremely complex so
that the final state of the system cannot be easily predicted from the
initial condition when several equilibrium states exist. In addition,
the structure of this basin of attraction depends whether the system
is described by the microcanonical or by the canonical ensemble
\cite{miller,crs}.

However, for $\Lambda>\Lambda_{c}$ the metastable phase completely
disapears and, in that case, the system {\it must} necessarily
collapse. This transition is associated to what has been traditionally
called the ``gravothermal catastrophe'' \cite{lbw} in the case of
classical point masses. For systems described by the Fermi-Dirac
statistics, the core ultimately ceases to shrink when it becomes
degenerate. In that case, the system falls on to the global entropy
maximum (points D) which is the true equilibrium state for these
systems. This global entropy maximum has a ``core-halo'' structure,
with a degenerate core and a non degenerate halo. This phase
transition is sometimes called a zeroth order phase transition
\cite{vega3} since it is associated with a discontinuous jump of
entropy (in the classical limit, the entropy of the condensed phase is
infinite). In fact, this does not correspond to a true phase
transition (in the usual sense), not even to an instability but simply
to the sudden disapearance of the ``gaseous'' phase. It has to be
noted that the degenerate nucleus resulting from this gravitational
collapse has a relatively important mass and a very small radius (for
$\mu=10^{5}$ and $\Lambda\simeq \Lambda_{c}$, we have typically
$M_{*}/M\simeq 0.22$ and $R_{*}/R\simeq 5\ 10^{-3}$).  This massive
nucleus (``fermion star'') can have important astrophysical
implications and, in the context of dark matter, may mimick the effect
of a black hole at the center of galaxies (see Sec. \ref{sec_astro}).

\subsection{Small values of the degeneracy parameter ($\mu=10^{3}$)}
\label{sec_small}

\begin{figure}[htbp]
\centerline{
\psfig{figure=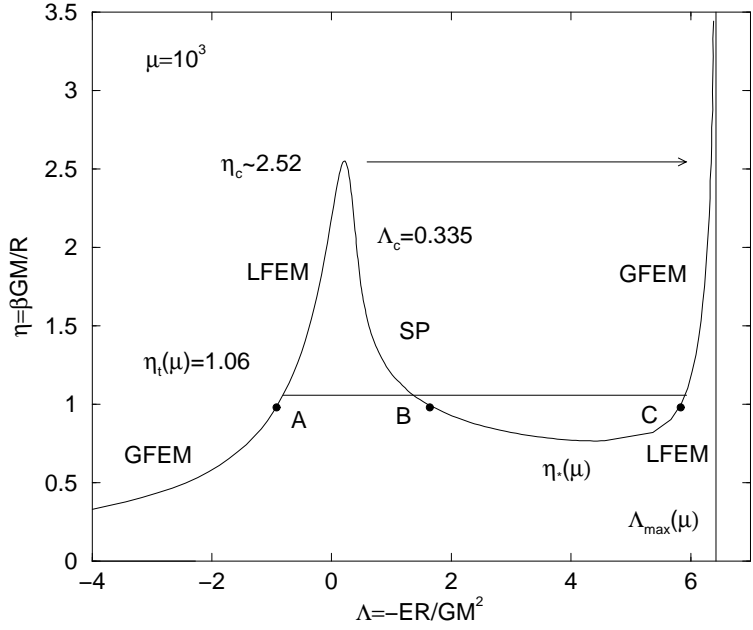,angle=0,height=8.5cm}}
\caption{Equilibrium phase diagram for Fermi-Dirac spheres with a degeneracy parameter  $\mu=10^{3}$. Points A form the ``gaseous'' phase. They are global maxima of free energy (GFEM) for $\eta<\eta_{t}$ and local maxima of free energy (LFEM) for $\eta>\eta_{t}$. The reverse is true for points C in the ``condensed'' phase. Points B are unstable saddle points (SP). This figure exhibits in particular a first order phase transition in the canonical ensemble. The equality of the free energy of the two phases ($J_{A}=J_{C}$) at the transition temperature $\eta_{t}$ implies the equality of the areas delimited by the
curve and the plateau,
i.e. $\int_{A}^{C}(\eta-\eta_{t})d\Lambda=0$. This is similar to
Maxwell's construction in the theory of the van der Waals gas.}
\label{fel}
\end{figure}

\begin{figure}[htbp]
\centerline{
\psfig{figure=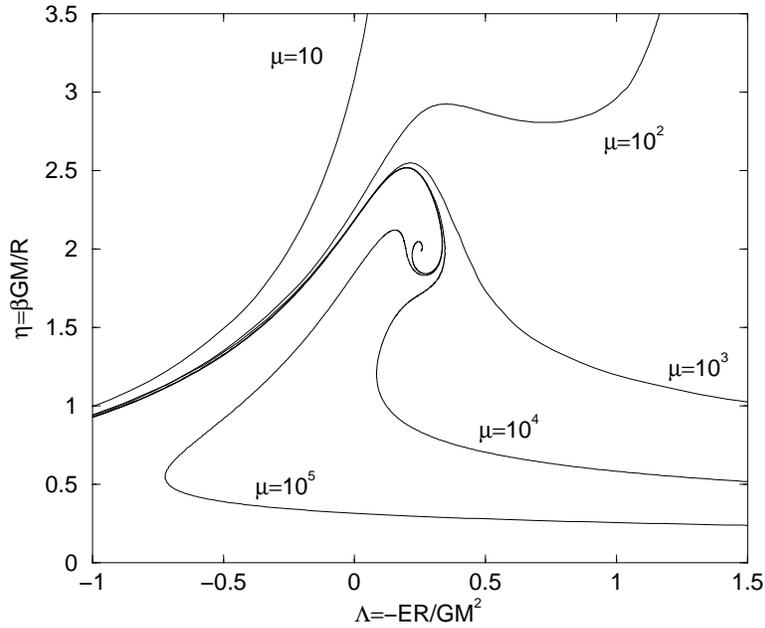,angle=0,height=8.5cm}}
\caption{Equilibrium phase
diagram for self-gravitating fermions with different values of the
degeneracy parameter $\mu$. The zeroth and first order phase transitions are
suppressed in the microcanonical ensemble for $\mu\lesssim 2600$ and
in the canonical ensemble for $\mu\lesssim 82.5$. For large values of
$\mu$, the curve makes several rotations before unwinding. The
criterion of Katz tells us that one mode of stability is lost each
time the curve rotates clockwise and regained as the curve rotates
anticlockwise. Therefore, only the upper and lower branches are
entropy maxima. For $\mu\rightarrow +\infty$ (classical limit), the
curve winds up indefinitely and tends to the spiral of Fig. \ref{etalambda}
as discussed in the text.}
\label{multimu}
\end{figure}

\begin{figure}[htbp]
\centerline{
\psfig{figure=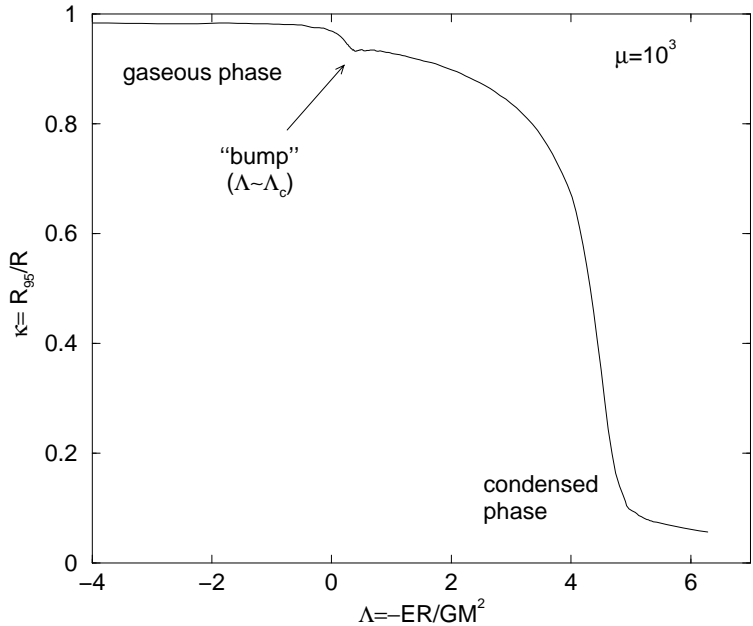,angle=0,height=8.5cm}}
\caption{Evolution of the order parameter with the energy for $\mu=10^{3}$. This figure illustrates the  ``clustering'' of the self-gravitating gas as energy is lowered. The presence of the ``bump'' at $\Lambda\sim\Lambda_{c}$ was previously noted by Follana \& Laliena \cite{laliena} in their model. }
\label{kappa}
\end{figure}

When the degeneracy parameter is sufficiently small, there exists only one
critical point of entropy for each value of energy (see
Fig. \ref{fel}) and it is a global entropy maximum. Therefore, a
sufficiently strong degeneracy suppresses the phase transitions in the
microcanonical ensemble, including the ``gravothermal catastrophe''
(see Fig. \ref{multimu}). For high energies (small values of
$\Lambda$) the solutions almost coincide with the classical isothermal
spheres. When the energy is lowered (large values of $\Lambda$) the
solutions take a ``core-halo'' structure with a partially degenerate
nucleus surrounded by a dilute Maxwellian atmosphere.  It is now
possible to overcome the critical energy $\Lambda_{c}=0.335$ and the
critical density contrast ${\cal R}=709$ found by Antonov
\cite{antonov} for classical particles. In that region, the specific heat
is negative, which is allowed in the microcanonical ensemble. As
energy decreases further, more and more mass is concentrated into the
nucleus (which becomes more and more degenerate) until a minimum
accessible energy, corresponding to $\Lambda_{max}(\mu)$, at which the
nucleus contains all the mass. In that case, the atmosphere has been
``swallowed'' and the system has the same structure as a cold white
dwarf star
\cite{chandra}. This is a relatively singular limit since the density
drops to zero at a finite radius whereas for partially degenerate
systems, the density decays like $r^{-2}$ at large distances. We can
study the formation of this compact object by defining an order
parameter $\kappa=R_{95}/R$, where $R_{95}$ is the radius of the
sphere which contains $95\%$ of the mass \cite{laliena}. This
parameter is ploted as a function of energy in Fig. \ref{kappa} and
the diagram is similar to the one obtained by Follana \& Laliena
\cite{laliena} with a different regularization of the potential at the
origin. For high energies, the density varies smoothly with the
distance and $\kappa\sim 1$. For low energies, the system
spontaneously forms a dense core containing more and more mass so that
$\kappa\rightarrow \kappa_{c}\ll 1$ (with $\kappa_{c}\simeq
6.678/\mu^{2/3}$, estimated from the mass-radius relation (\ref{c5})
of a completely degenerate nucleus). We observe that the order
parameter varies rapidly in the region of negative specific heats but
remains continuous. According to Cerruti-Sola {\it et al.} 
\cite{sola}, this is the mark of a second order phase transition at
$\Lambda=\Lambda_{0}$ (corresponding to the point of minimum
temperature $\eta_{c}$). At that point, the specific heat is infinite
and turns from positive to negative; more precisely, $C=dE/dT$
diverges like $(\Lambda-\Lambda_{0})^{-1} $ and
$\pm(\eta_{c}-\eta)^{-1/2}$ \cite{chaviso}. In
fact, this ``second order phase transition'' is not really a
phase transition; it just corresponds to the ``clustering'' of the
self-gravitating gas as its energy is progressively reduced.

\begin{figure}[htbp]
\centerline{
\psfig{figure=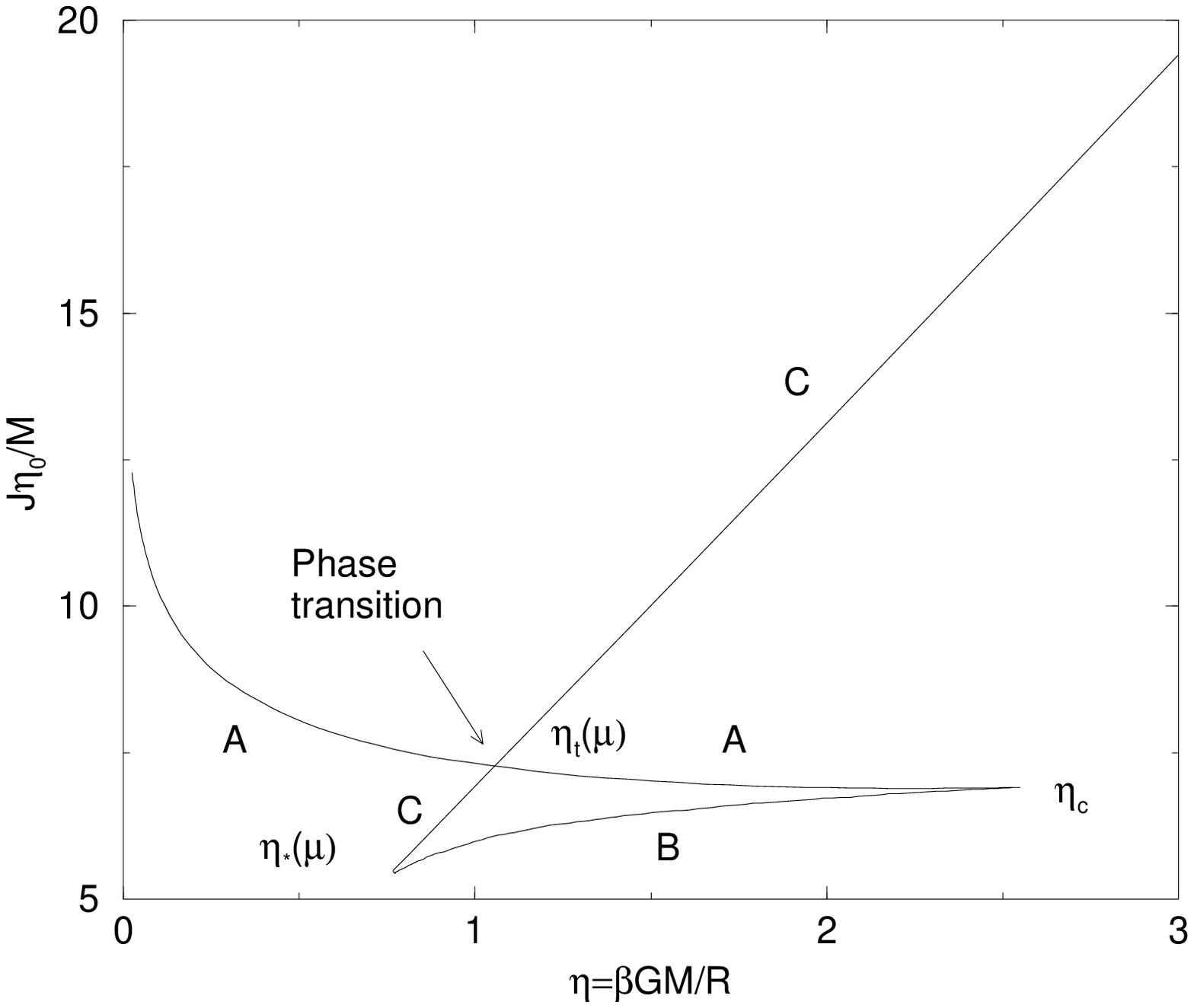,angle=0,height=8.5cm}}
\caption{Free energy of each phase versus inverse temperature for $\mu=10^{3}$. The interpreatation is the same as in Fig. \ref{SL5}.}
\label{fetaJ}
\end{figure}

If we now consider the canonical situation, we see that several
solutions exist at the same temperature. A first order phase
transition occurs at $\eta_{t}(\mu)$ an separates a ``gaseous'' phase
from a ``condensed'' phase.  The interpretation is the same as in the
microcanonical ensemble. The energy is discontinuous at the transition
so that a large amount of latent heat is released. The specific heat
is also discontinuous but remains positive.  For $\eta>\eta_{t}(\mu)$
the ``gaseous'' states are {\it metastable} but they probably are
physical. This metastable branch completely disapears at
$\eta=\eta_{c}$ and the system undergoes an ``isothermal
collapse''. This phase transition is more radical than the previous
one since it is marked by the disapearance of the metastable
phase. For self-gravitating fermions the ``isothermal collapse'' ends
up on a compact state with a core-halo structure (for $\mu=10^{3}$,
$M_{*}\simeq M$ and $R_{*}\simeq 6.7\ 10^{-2}R$ at $\eta\sim
\eta_{c}$).  Since the free energy is discontinuous (see
Fig. \ref{fetaJ}), this could be called a zeroth order phase
transition \cite{vega3}. It has been shown in Ref. \cite{chaviso} that
the point of minimum temperature $\eta_{c}$ coincides with the Jeans
instability criterion. More precisely, the condition of instability of
an isothermal gas sphere with respect to linear perturbations of the
Navier-Stokes equation can be written $R> (\eta_{c}/3)^{1/2}L_{J}\sim
L_{J}$, where $L_{J}$ is the Jeans length.  Since $\eta_{t}<\eta_{c}$,
a phase transition (corresponding to a nonlinear evolution of the
system) can occur at scales much smaller than the Jeans scale
($R=(\eta_{t}/3)^{1/2} L_{J}\ll L_{J}$). This might explain the
formation of smaller objects than usually achieved with the ordinary
Jeans instability. This idea has been developed by Stahl {\it et al.} 
\cite{stahl} in relation with planet formation. However, this picture
is limited by the realization that the evolution of the system depends
on a complicated notion of ``basin of attraction'' which is difficult
to describe in detail (see Sec. \ref{sec_large}). In particular, it is
not clear whether the true phase transition will occur at $\eta_{t}$.

\section{Tricritical points}
\label{sec_tricritical}

We have indicated in the preceding section that the phase transition in the
microcanonical ensemble disapears at a critical degeneracy parameter
$\mu_{MTP}\simeq 2600$ and that the phase transition in the
canonical ensemble disapears for $\mu>\mu_{CTP}\simeq 82.5$. These
points at which the $\Lambda-\eta$ curve presents an inflexion  
are called {\it tricritical points} (TP) in the langage of phase transitions.

\begin{figure}[htbp]
\centerline{
\psfig{figure=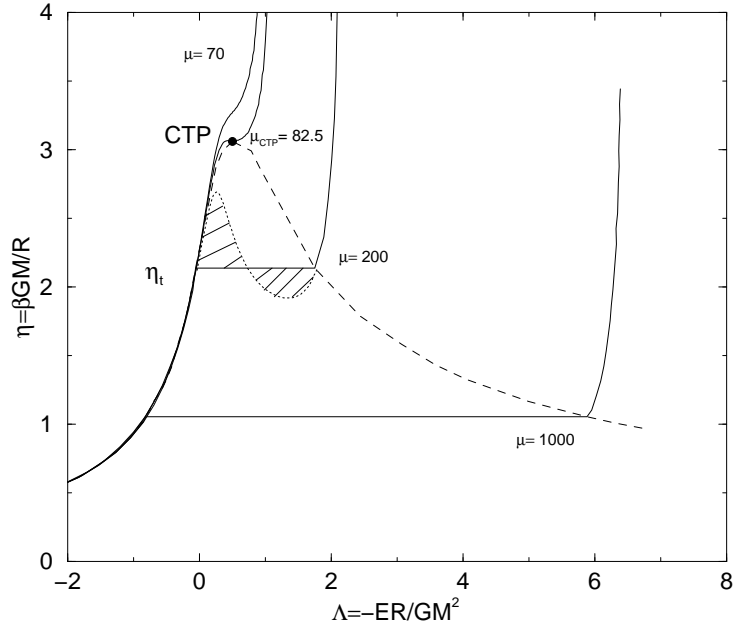,angle=0,height=8.5cm}}
\caption{Enlargement of the phase diagram near the
tricritical point in the canonical ensemble. The Maxwell construction
determining the transition temperature $\eta_{t}(\mu)$ is done
explicitly (dashed areas). For $\mu_{CTP}=82.5$ the Maxwell plateau
disapears and the $\Lambda-\eta$ curve presents an inflexion point at
$\Lambda_{CTP}\simeq 0.5$, $\eta_{CTP}\simeq 3.06$. At that point, the
specific heat becomes infinite and the transition is second
order. This diagram is remarkably similar to the liquid/gas transition
for an ordinary fluid.}
\label{tricritique}
\end{figure}

In Fig. \ref{tricritique}, we have enlarged the phase diagram near the
tricritical point in the canonical ensemble. It is located at
$\Lambda_{CTP}\simeq 0.5$ and $\eta_{CTP}\simeq 3.06$. In the canonical
ensemble, the oscillations of the $\Lambda-\eta$ curve for
$\mu>\mu_{CTP}$ are replaced by a horizontal Maxwell plateau
connecting the gaseous phase (left) to the condensed phase
(right). This characterizes a canonical first order phase transition
at $\eta_{t}(\mu)$. This diagram exhibits a close analogy with the
classical gas/liquid transition, the liquid phase beeing the
counterpart of the (gravitational) Fermi condensate. At the
tricritical point $CTP$, the plateau disapears and the specific heats
diverges like $C\sim (\Lambda-\Lambda_{c})^{-2}\sim
|\eta-\eta_{c}|^{-2/3}$. This is similar to the $\lambda$-transition
in $^{4}H_{e}$. Therefore, the first order phase transition becomes
second order at the canonical tricritical point.

\begin{figure}[htbp]
\centerline{
\psfig{figure=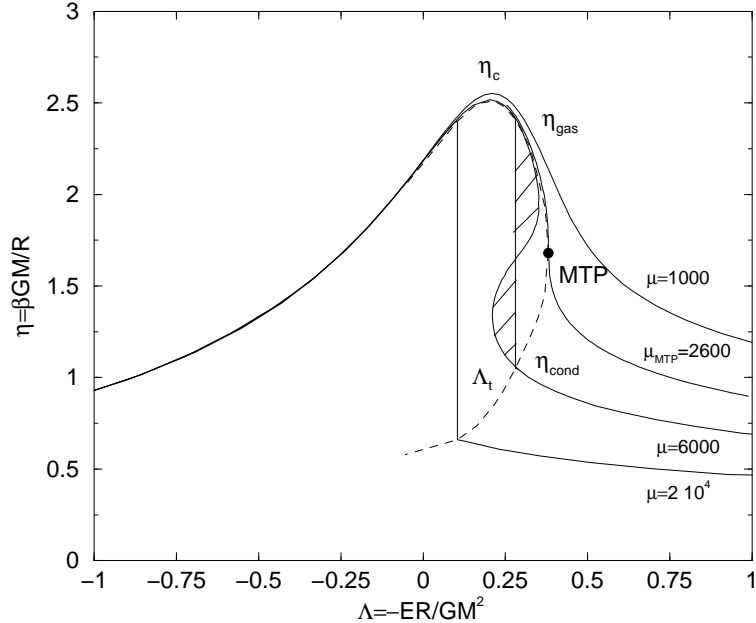,angle=0,height=8.5cm}}
\caption{Enlargement of the phase diagram near the
tricritical point in the microcanonical ensemble. The Maxwell
construction determining the transition energy $\Lambda_{t}(\mu)$ is
done explicitly (dashed areas). For $\mu_{MTP}=2600$ the Maxwell
plateau disapears and the $\Lambda-\eta$ curve presents an inflexion
point at $\Lambda_{MTP}\simeq 0.38$, $\eta_{MTP}\simeq 1.68$. We have
indicated different characteristic temperatures, as described in the
text.}
\label{tricritiquemicro}
\end{figure}
 
The gravitational Fermi gas diagram is nevertheless more complex than
the liquid/gas diagram because it presents {\it another} tricritical
point $MTP$ in the microcanonical ensemble. In
Fig. \ref{tricritiquemicro}, we have enlarged the phase diagram near
this tricritical point. It is located at $\Lambda_{MTP}\simeq 0.38$,
$\eta_{MTP}\simeq 1.68$. The interpretation is the same as in the
canonical ensemble except that the plateau is now vertical as it
corresponds to a discontinuity of temperature at a transition energy
$\Lambda_{t}(\mu)$ (microcanonical first order phase transition). We
have denoted by $\eta_{gas}$ and $\eta_{cond}$ the values of the
inverse temperature of the two phases at the transition energy. At the
tricritical point $\eta_{gas}=\eta_{cond}$.  We have also indicated in
the figure the point of minimum temperature $\eta_{c}$ at which the
specific heat diverges (second order microcanonical phase transition).

\begin{figure}[htbp]
\centerline{
\psfig{figure=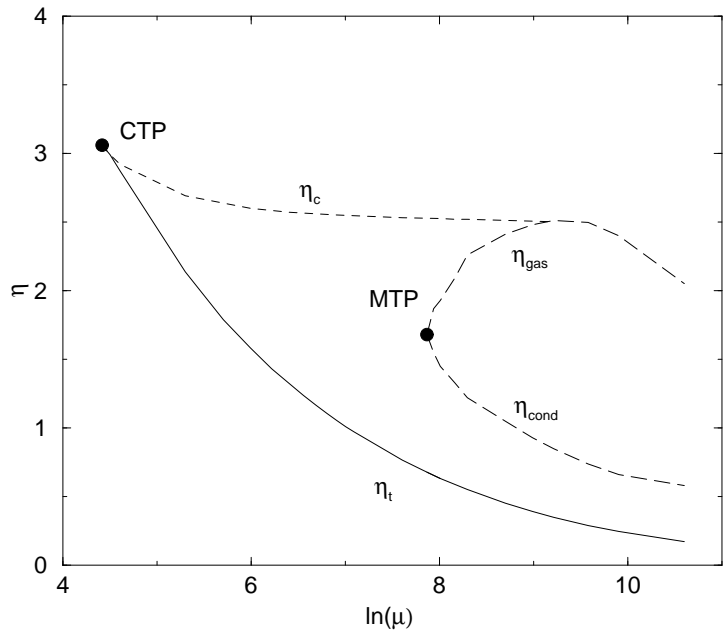,angle=0,height=8.5cm}}
\caption{Phase diagram of the self-gravitating Fermi gas in the $\mu-\eta$ plane. The solid line gives the transition temperature at the canonical first order phase transition. The dashed lines give the temperatures of the two phases (gas and condensate) at the microcanonical first order phase transition. The dotted line gives the critical temperature $\eta_{c}$ at the microcanonical second order phase transition. CTP and MTP are the tricritical points in the two ensembles.}
\label{diagRuffo}
\end{figure}

In Fig. \ref{diagRuffo}, we have ploted the values of the transition
temperature $\eta_{t}$ in the canonical ensemble and the values of the
characteristic temperatures $\eta_{gas}$ and $\eta_{cond}$ in the
microcanonical ensemble as a function of the degeneracy parameter
$\mu$. These curves characterize first order phase transitions in the
canonical and microcanonical ensembles. We have also represented the
minimum temperature $\eta_{c}$ marking the onset of a microcanonical
second order phase transition. At the canonical tricritical point CTP,
this branch connects the $\eta_{t}$-branch. This is consistent with
the appearance of an {\it isolated} second order phase transition in
the canonical ensemble at CTP. This diagram shares some similitudes
with the one obtained by Barr\'e {\it et al.} \cite{barre} in their
analysis of the Blume-Emery-Griffiths (BEG) model with infinite range
interactions. This system also displays an inequivalence of ensembles,
regions of negative specific heats and two tricritical points (one in
each ensemble). This analogy suggests that these properties are
related to the long-range nature of the interactions more than to the
details of the model. This implies a kind of universality for this
type of systems. However, there are also noticable differences between
the two models. In particular, in the BEG model, second order phase
transitions are characterized by a discontinuity of the specific heat
(angular point) while in our model of self-gravitating fermions the
specific heat is always continuous except at the critical point
$\eta_{c}$ at which it diverges (this is our definition of a second
order phase transition). Therefore, there still exists a second order
phase transition above the canonical tricritical point in the BEG
model but not in the self-gravitating Fermi gas. Another consequence
of the continuity of the specific heats in our model is that the
second order critical line $\eta_{c}-\mu$ (dotted line) does not
connect the first order critical lines (dashed lines) at the
microcanonical tricritical point MTP but slightly after, unlike in the
BEG model \cite{barre}.

\section{A simple analytical model for self-gravitating fermions}
\label{sec_analytical}

The previous study has revealed that self-gravitating fermions can
undergo a phase transition from a ``gaseous'' phase to a ``condensed''
phase. We shall now propose a simple analytical model to describe this
phase transition more conveniently. As we shall see, our model can
reproduce remarkably well the essential features of the equilibrium
phase diagram and it can be used to determine the dependance of the
critical parameters with the degeneracy parameter $\mu$.

\subsection{The ``gaseous'' phase}
\label{sec_gas}

The ``gaseous'' phase can be represented by a homogeneous distribution of particles with a Maxwellian distribution of velocities 
\begin{equation}
f={\rho\over (2\pi T)^{3/2}}e^{-{v^{2}\over 2T}}.
\label{g1}
\end{equation} 
The relation between the energy and the temperature is given by
\begin{equation}
E={3\over 2}MT-{3GM^{2}\over 5R},
\label{g2}
\end{equation} 
and the entropy by
\begin{equation}
{\eta_{0}S\over M}={3\over 2}\ln (2\pi T)-\ln\biggl ({3M\over 4\pi R^{3}}\biggr )+\ln \eta_{0}+{3\over 2}.
\label{g3}
\end{equation} 
Introducing the normalized energy $\Lambda$ and the normalized temperature $\eta$ defined in section \ref{sec_para}, we can rewrite the previous equations in the form
\begin{equation}
\Lambda={3\over 5}-{3\over 2\eta},
\label{g4}
\end{equation} 
\begin{equation}
{\eta_{0}S\over M}=-{3\over 2}\ln\eta+\ln\mu+{3\over 2}
+{1\over 2}\ln\pi-\ln6.  
\label{g5}
\end{equation} 
These equations correctly describe the gaseous phase for high energies
and high temperatures (i.e. low density contrasts). Of course, it can
not reproduce the spiral behaviour of the classical phase diagram
which is an intrinsic property of the Emden equation. An analytical
expression for this spiral has been given in Ref. \cite{chaviso} in the 
asymptotic limit of high density contrasts.

\subsection{The ``condensed'' phase}
\label{sec_cp}

For the ``condensed'' phase, we shall improve the core-halo model proposed by Chavanis \& Sommeria \cite{cs}. We assume that the ``core'' is completely degenerate and we denote by $M_{*}$, $R_{*}$ and $E_{*}$ its mass, radius and energy. In that limit, the distribution function is a step function: $f=\eta_{0}$ for $v\le v_{max}$ and $f=0$ for $v>v_{max}$. Of course, for a self-gravitating system, the maximum velocity $v_{max}$ is a function of the position. For this simple distribution function, the pressure and the density are given by
\begin{equation}
p={1\over 3}\int fv^{2}d^{3}{\bf v}=
{4\pi \eta_{0}\over 3}{v_{max}^{5}\over 5}, 
\label{c1}
\end{equation}  
\begin{equation}
\rho=\int f d^{3}{\bf v}=4\pi\eta_{0}{v_{max}^{3}\over 3}. 
\label{c2}
\end{equation}
Eliminating the velocity between these two relations, we find that the equation of state of a completely degenerate system is that of a polytrope with index  $\gamma=5/3$ (or $n=3/2$),
\begin{equation}
p=K\rho^{5/3},\qquad K={1\over 5}\biggl ({3\over 4\pi\eta_{0}}\biggr )^{2/3}.
\label{c3}
\end{equation} 
These results are of course well-known from the theory of white dwarf stars \cite{chandra} and they are repeated only in order to determine the constant $K$ in the present context.  Now, from the theory of polytropic spheres \cite{chandra}, the mass-radius relation is given in the general case by
\begin{equation}
K=N_{n}GM^{(n-1)/n}R^{(3-n)/n},
\label{c4}
\end{equation} 
where $N_{n}$ is a constant depending on the index of the polytrope. For $n=3/2$, one has $N_{3/2}=0.42422...$. Therefore, the relation between the mass and the radius of our degenerate nucleus is
\begin{equation}
M_{*}R_{*}^{3}={\chi\over \eta_{0}^{2}G^{3}}, \quad {\rm with}\quad  \chi\simeq 5.9723\ 10^{-3}.
\label{c5}
\end{equation} 
On the other hand, its energy is given by \cite{chandra}
\begin{equation}
E_{*}=-{3\over 7}{GM_{*}^{2}\over R_{*}},
\label{c6}
\end{equation} 
and its entropy is equal to zero since the distribution function is unmixed ($f=\eta_{0}$).

By shrinking, the nucleus releases an enormous amount of energy which heats the envelope. The envelope behaves therefore like an ordinary gas maintained by the walls of the box so that its density is approximately uniform. Its energy and entropy are therefore given by
\begin{equation}
E_{halo}={3\over 2}(M-M_{*})T-{3GM_{*}(M-M_{*})\over 2R}-{3G(M-M_{*})^{2}\over 5R},
\label{c7}
\end{equation} 
\begin{equation}
\eta_{0}S=(M-M_{*})\biggl \lbrack {3\over 2}\ln (2\pi T)-\ln(M-M_{*})+\ln V+\ln\eta_{0}+{3\over 2}\biggr\rbrack.
\label{c8}
\end{equation} 
Contrary to our previous paper \cite{cs}, we have not neglected the potential energy of the envelope as compared to its thermal energy. This sensibly improves the agreement  with the full numerical solution. However, we have still assumed that the core is much smaller than the halo, so that $V={4\over 3}\pi R^{3}$ represents the total volume of the system.  For calculating the potential energy, we have used the formula
\begin{equation}
W=-4\pi G\int \rho M(r) r dr,
\label{c9}
\end{equation} 
where $M(r)$ is the mass contained within the sphere of radius $r$. This formula is valid for an arbitrary spherically symmetrical distribution of matter \cite{bt}.

Adding Eqs. (\ref{c6}) (\ref{c7}) and expressing the radius of the core as a function of its mass, using Eq. (\ref{c5}), the total energy of the system is given by
\begin{equation}
E=-{3\over 7}{\eta_{0}^{2/3}G^{2}\over \chi^{1/3}}M_{*}^{7/3}+{3\over 2}(M-M_{*})T-{3GM_{*}(M-M_{*})\over 2R}-{3G(M-M_{*})^{2}\over 5R}.
\label{c10}
\end{equation}  
For a given value of energy (microcanonical ensemble), this relation determines the temperature $T$ as a function of the core mass. Therefore, the entropy (\ref{c8}) is a function of $M_{*}$ alone. The mass of the nucleus at equilibrium  is determined by maximizing the entropy with respect to $M_{*}$. After simplification, the condition $dS/dM_{*}=0$ is found to be equivalent to
\begin{eqnarray}
\ln(M-M_{*})+{3G\over 2RT}(M-2M_{*})-{6G\over 5RT}(M-M_{*})-{3\over 2}\ln T+{G^{2}\eta_{0}^{2/3}\over \chi^{1/3}}{M_{*}^{4/3}\over T}\nonumber\\
=\ln\eta_{0}+\ln V-1+{3\over 2}\ln (2\pi).
\label{c11}
\end{eqnarray}
We obtain the same relation by maximizing the free energy $J$ at fixed mass and temperature. Eqs. (\ref{c10})-(\ref{c11}) completely determine the equilibrium phase diagram of self-gravitating fermions in the framework of our analytical model. For convenience, we shall  re-express these equations in a dimensionless form. To that purpose, we introduce the fraction of mass $\alpha$ contained in the core such that $M_{*}=\alpha M$. In terms of $\alpha$, the radius of the core is $R_{*}/R=6.678\ \alpha^{-1/3}\mu^{-2/3}$ (see Eq. (\ref{c5})). On the other hand, we set
\begin{eqnarray}
\lambda={1\over (512\pi^{4}\chi)^{1/3}}=0.149756..., \qquad C={1\over 2}\ln\pi-\ln 6-1=-2.21939...
\label{c12}
\end{eqnarray}
Introducing furthermore the dimensionless energy $\Lambda$ and the dimensionless temperature $\eta$ defined in Sec. \ref{sec_para}, the equations of the problem become
\begin{eqnarray}
\Lambda={3\over 7}\lambda\mu^{2/3}\alpha^{7/3}-{3\over 2}(1-\alpha){1\over\eta}+{3\over 2}\alpha(1-\alpha)+{3\over 5}(1-\alpha)^{2},
\label{c13}
\end{eqnarray}
\begin{eqnarray}
\ln(1-\alpha)+{3\over 2}\ln\eta+\lambda\mu^{2/3}\alpha^{4/3}\eta+{9\over 5}\eta \biggl ({1\over 6}-\alpha\biggr )=\ln\mu+C,
\label{c14}
\end{eqnarray}
\begin{eqnarray}
{\eta_{0}S\over M}=(1-\alpha)\biggl\lbrack -{3\over 2}\ln\eta-\ln(1-\alpha)+\ln\mu+C+{5\over 2}\biggr\rbrack,
\label{c15}
\end{eqnarray}
\begin{eqnarray}
{\eta_{0}J\over M}={\eta_{0}S\over M}+\eta\Lambda.
\label{c16}
\end{eqnarray}

\begin{figure}[htbp]
\centerline{
\psfig{figure=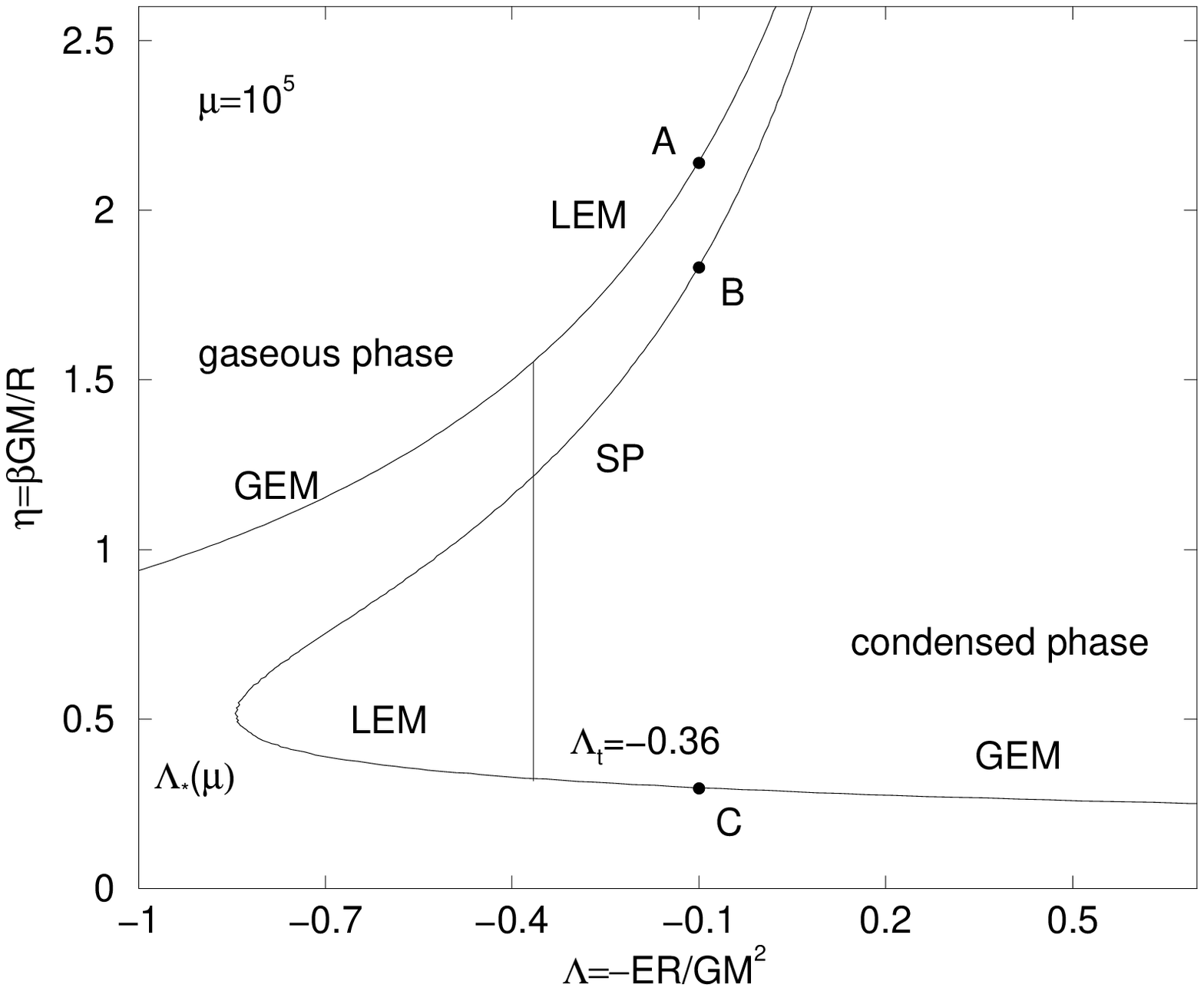,angle=0,height=8.5cm}}
\caption{Equilibrium phase diagram obtained from our analytical model $(\mu=10^{5})$. It compares relatively well with the full numerical solution reported in Fig. \ref{le5}}.
\label{model5}
\end{figure}

\begin{figure}[htbp]
\centerline{
\psfig{figure=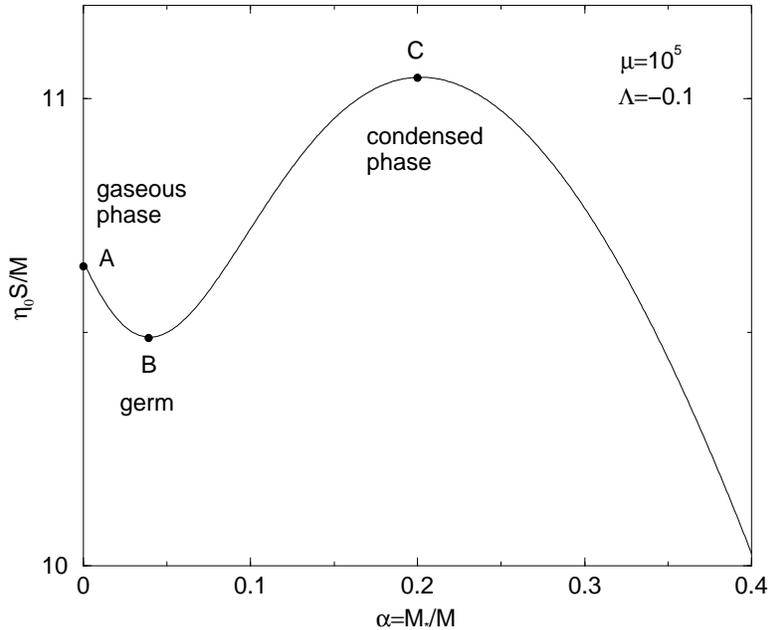,angle=0,height=8.5cm}}
\caption{Entropy as a function of the mass of the nucleus for $\mu=10^{5}$ and $\Lambda=-0.1>\Lambda_{t}$. This curve is obtained by using the analytical formulae (\ref{c13})(\ref{c14})(\ref{c15}). For $\Lambda>\Lambda_{t}$, the ``condensed'' state (point C) is a global entropy maximum and the ``gaseous'' state (point A) a local entropy maximum. The solution with the ``germ'' (point B) is an entropy minimum.  The fraction of mass contained in the nucleus is relatively
large for $\mu=10^{5}$ (see Fig. \ref{Lalp5}) but it decreases as
the classical limit is approached. As $\mu\rightarrow +\infty$,
$\alpha_{cond}\rightarrow 0$, $\alpha_{germ}\rightarrow 0$ with
$\alpha_{cond}\gg\alpha_{germ}$.}
\label{entropie}
\end{figure}

\begin{figure}[htbp]
\centerline{
\psfig{figure=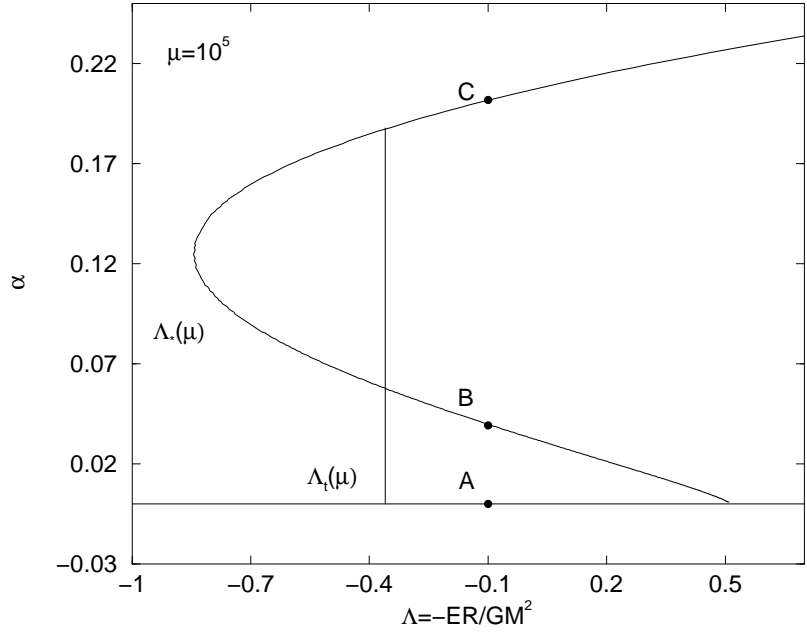,angle=0,height=8.5cm}}
\caption{Evolution of the fraction of mass contained in the nucleus as a function of energy for a degeneracy parameter $\mu=10^{5}$.}
\label{Lalp5}
\end{figure}

\begin{figure}[htbp]
\centerline{
\psfig{figure=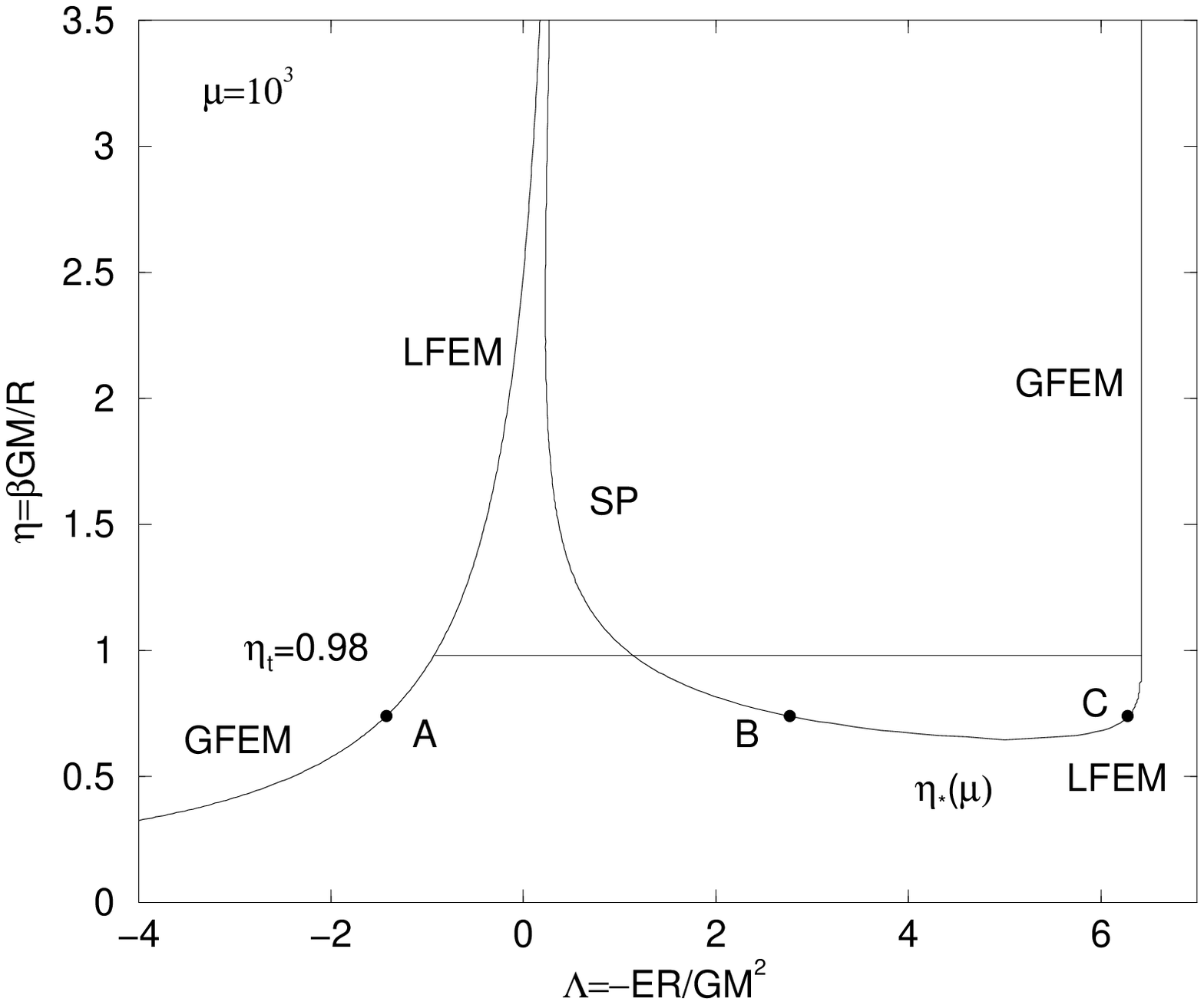,angle=0,height=8.5cm}}
\caption{Equilibrium phase diagram obtained from our analytical model $(\mu=10^{3})$. It compares relatively well with the full numerical solution reported in Fig. \ref{fel}.}
\label{model3}
\end{figure}

\begin{figure}[htbp]
\centerline{
\psfig{figure=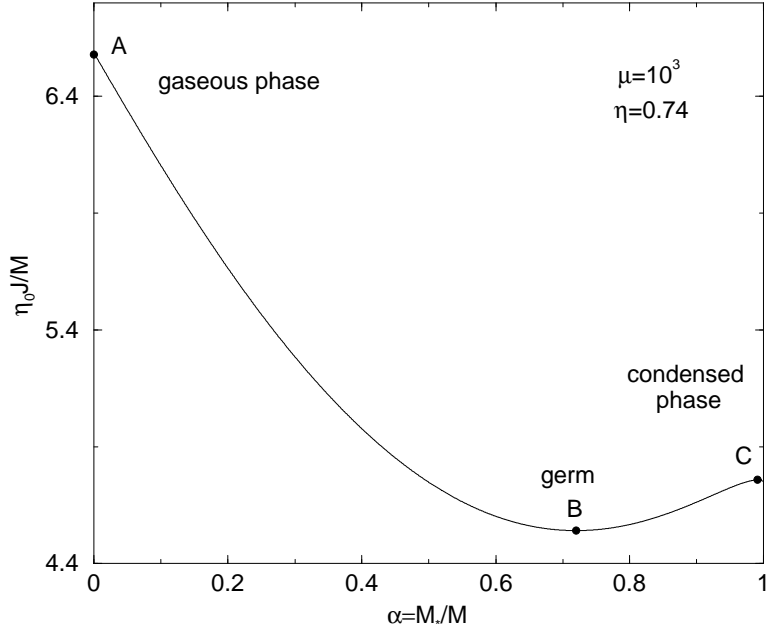,angle=0,height=8.5cm}}
\caption{Free energy as a function of the mass of the nucleus for $\mu=10^{3}$ and $\eta=0.74<\eta_{t}$. For $\eta<\eta_{t}$, the ``gaseous'' state is a global maximum of free energy and the ``condensed'' state is a local maximum of free energy. The solution with the ``germ''  is a minimum of free energy. The fraction of mass contained in the condensate is close to one (see Fig. \ref{Lalp3}). In the limit $\mu\rightarrow \infty$, $\alpha_{cond}\rightarrow 1$ and $\alpha_{germ}\rightarrow 0$.}
\label{elibre}
\end{figure}

\begin{figure}[htbp]
\centerline{
\psfig{figure=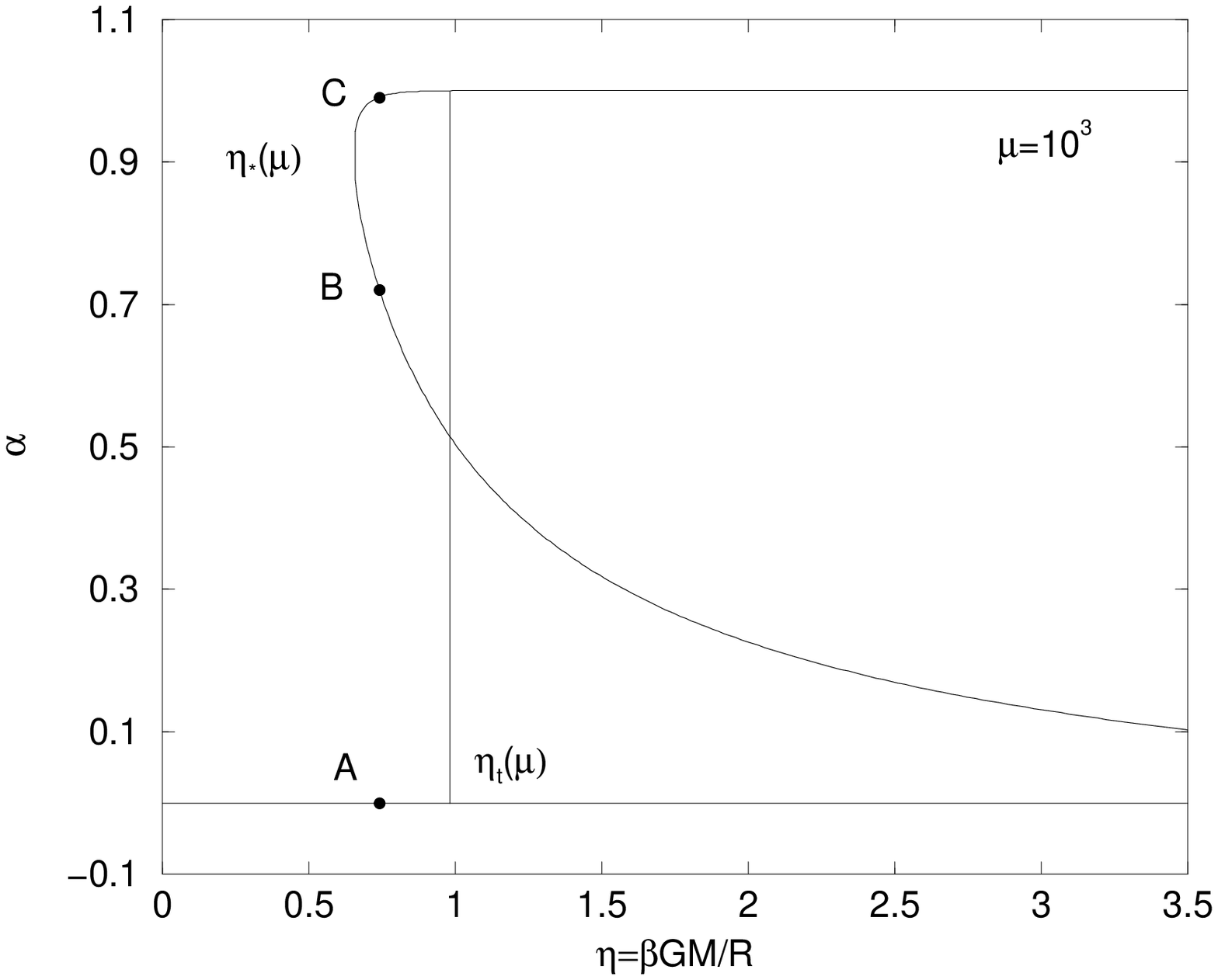,angle=0,height=8.5cm}}
\caption{Evolution of the fraction of mass contained in the nucleus as a function of the inverse temperature for a degeneracy parameter $\mu=10^{3}$.}
\label{Lalp3}
\end{figure}

The equilibrium phase diagram is represented in Fig. \ref{model5} for
a degeneracy parameter $\mu=10^5$. It provides a fairly good agreement
with the full numerical solution of Sec. \ref{sec_large} (see
Fig. \ref{le5}). Of course, we cannot expect to reproduce exactly the
numerical results in view of the approximations made in our analytical
model. In particular, except for very low energies, the core is only
partially degenerate and this is responsible for the quantitative
discrepencies observed between the two diagrams. However, the
qualitative behaviour is the same and this is essentially what was
attempted by our analytical approach. In particular, we recover the
three types of solutions previously studied. The solutions on the
upper branch (points A) form the ``gaseous'' phase. They can also be
considered as a particular limit of the core-halo model with
$\alpha=0$.  The solutions on the lower branch (points C) form the
``condensed'' phase and the solutions on the intermediate branch
(points B) are similar to the ``gaseous'' states (points A) except
that they possess a small central nucleus (a ``germ''). To determine
the stability of these solutions we have ploted in Fig. \ref{entropie}
their entropy as a function of the core mass $\alpha$ for a given
energy. The curve $S(\alpha)$ has the usual ``$W$-shape''
characteristic of phase transitions.  The ``gaseous'' state
($\alpha=0$) can be considered as an entropy maximum although it does
not correspond to the condition $dS/d\alpha=0$. The other entropy maximum
corresponds to the ``condensed'' state (point C) and the
entropy minimum to the solution with the germ (point
B). For $\Lambda>\Lambda_{t}$, the ``condensed'' states $C$ are global
entropy maxima and the ``gaseous'' states $A$ are local entropy maxima
(the reverse is true for $\Lambda<\Lambda_{t}$). However, to pass from
$A$ to $C$, we have to cross an entropic barrier constituted by the
solution $B$. This requires that the entropy {\it decreases} which is
not possible for an isolated system. Therefore, depending whether the
initial fraction of mass $\alpha_{0}$ is smaller or larger than
$\alpha_{germ}$, the system can either relax towards the ``gaseous''
state or collapse towards the ``condensed'' state. Of course, this
argument assumes that $\alpha$ is the only degree of freedom in the
system, which is clearly an idealization. As mentioned previously, the
real ``basin of attraction'' is much more complicated.

The equilibrium phase diagram corresponding to a degeneracy parameter
$\mu=10^3$ is represented in Fig. \ref{model3} and it compares
relatively well with the full numerical solution (see
Fig. \ref{fel}). The minimum energy, corresponding to
$\Lambda_{max}(\mu)$, is reached when all the mass is in the
degenerate core at zero temperature ($\alpha\rightarrow 1$,
$\eta\rightarrow +\infty$). Eq. (\ref{c13}) then yields
\begin{eqnarray}
\Lambda_{max}(\mu)={3\over 7}\lambda\mu^{2/3}.
\label{c17}
\end{eqnarray} 
For $\Lambda\rightarrow \Lambda_{max}(\mu)$ and for sufficiently large $\mu$, one has
\begin{eqnarray}
1-\alpha\sim e^{-\lambda\mu^{2/3}\eta}, \qquad \Lambda_{max}-\Lambda\sim \lambda\mu^{2/3}e^{-\lambda\mu^{2/3}\eta}.
\label{c18}
\end{eqnarray} 
This shows that the atmosphere is swallowed exponentially rapidly when
we approach the minimum energy. Fig. \ref{model3} displays a clear
phase transition in the canonical ensemble. The value of the
temperature of transition is very close to the exact value found in
the numerical approach. In the present case, we can safely consider
that the core of the condensate is completely degenerate so that the
quantitative agreement with the exact solution is better than in the
microcanonical ensemble. Once again, the stability of the equilibrium
states can be determined by considering the variation of the free
energy with the core mass (see Fig. \ref{elibre}). For $\eta<\eta_{t}$
(resp. $\eta>\eta_{t}$), the ``gaseous'' state is a global
(resp. local) maximum of free energy and the ``condensed'' state a
local (resp. global) one.  The description of the phase transition is
the same as in the microcanonical ensemble.

\subsection{Scaling laws in the limit $\mu\rightarrow +\infty$ }
\label{sec_scaling}

We shall now determine the behaviour of the critical parameters as $\mu\rightarrow +\infty$ (classical limit). In the microcanonical ensemble, the phase transition occurs close to the maximum energy $E_{*}$ of the condensed phase, corresponding to $\Lambda_{*}(\mu)$. Using Eqs. (\ref{c13}) (\ref{c14}), we find that the condition of energy maximum $d\Lambda=0$ is equivalent to 
\begin{eqnarray}
\biggl\lbrack {3\over 2}+\lambda\mu^{2/3}\alpha^{4/3}\eta+{9\over 5}\eta\biggl ({1\over 6}-\alpha\biggr )\biggr\rbrack^{2}={3\over 2}(1-\alpha)\biggl ({4\over 3}\lambda\mu^{2/3}\alpha^{1/3}\eta-{1\over 1-\alpha}-{9\over 5}\eta\biggr ).
\label{s1}
\end{eqnarray}
This equation, together with Eqs. (\ref{c13}) (\ref{c14}),  determines the function $\Lambda_{*}(\mu)$. Now, in the limit $\mu\rightarrow +\infty$, the fraction of mass contained in the nucleus goes to zero while the temperature increases. Taking the limit $\mu\rightarrow +\infty$, $\alpha\rightarrow 0$ and $\eta\rightarrow 0$ in Eq. (\ref{s1}), we obtain the relation 
\begin{eqnarray}
\lambda\mu^{2/3}\alpha^{7/3}\eta=2.
\label{s2}
\end{eqnarray} 
Considering the limiting form of  Eqs. (\ref{c13}) (\ref{c14}) in the same approximation, we find
\begin{eqnarray}
\Lambda={3\over 7}\lambda\mu^{2/3}\alpha^{7/3}-{3\over 2\eta},
\label{s3}
\end{eqnarray} 
\begin{eqnarray}
{3\over 2}\ln\eta+\lambda\mu^{2/3}\alpha^{4/3}\eta=\ln\mu.
\label{s4}
\end{eqnarray} 
Therefore, the thermodynamical parameters at the point of maximum
energy behave with the degeneracy parameter like
\begin{eqnarray}
\alpha_{*}\sim {1\over\ln\mu},\quad \eta_{*}\sim {(\ln\mu)^{7/3}\over \mu^{2/3}},\quad \Lambda_{*}\sim  -{\mu^{2/3}\over (\ln\mu)^{7/3}}. 
\label{s5}
\end{eqnarray}  

The transition point $\Lambda_{t}(\mu)$ is determined by equating the
entropy of the two phases. This yields
\begin{eqnarray}
(1-\alpha)\biggl\lbrack -{3\over 2}\ln\eta_{cond}-\ln(1-\alpha)+\ln\mu+C+{5\over 2}\biggr\rbrack=
-{3\over 2}\ln\eta_{gas}+\ln\mu+C+{5\over 2},
\label{snew1}
\end{eqnarray} 
where $\eta_{cond}$ is the temperature of the condensed phase, $\eta_{gas}$ the temperature of the gaseous phase and $\alpha$ the mass contained in the condensate at the transition point. Considering the limiting form of Eqs. (\ref{g4}) (\ref{c13}) (\ref{c14}) when $\mu\rightarrow +\infty$, $\alpha\rightarrow 0$ and $\eta\rightarrow 0$, we find that
\begin{eqnarray}
\Lambda=-{3\over 2\eta_{gas}}={3\over 7}\lambda\mu^{2/3}\alpha^{7/3}-{3\over 2\eta_{cond}},
\label{wq1}
\end{eqnarray} 
\begin{eqnarray}
{3\over 2}\ln\eta_{cond}+\lambda\mu^{2/3}\alpha^{4/3}\eta_{cond}=\ln\mu.
\label{wq2}
\end{eqnarray} 
Solving for $\eta_{cond}$ and $\eta_{gas}$, we get
\begin{eqnarray}
\eta_{cond}\sim {2\ln\mu\over \lambda\mu^{2/3}\alpha^{4/3}},\qquad {1\over\eta_{gas}}\sim {\lambda\mu^{2/3}\alpha^{4/3}\over 2\ln\mu}\biggl (1-{4\over 7}\alpha\ln\mu\biggr ). 
\label{snew2}
\end{eqnarray} 
Substituting these results in Eq. (\ref{snew1}), we find that the mass contained in the condensate behaves like
\begin{eqnarray}
\alpha\sim {1\over \ln \mu}.
\label{snew3}
\end{eqnarray} 
Combining the foregoing relations, we find that the energy of transition is given by
\begin{eqnarray}
\Lambda_{t}\sim  -{\mu^{2/3}\over (\ln\mu)^{7/3}}.
\label{snew4}
\end{eqnarray} 
In the classical limit, $\Lambda_{t}\rightarrow -\infty$, so that the
``gaseous'' states are always metastable as previously discussed. The
temperature of the two phases at the transition point behave like
\begin{eqnarray}
\eta_{gas}\sim \eta_{cond}\sim {(\ln\mu)^{7/3}\over \mu^{2/3}}.
\label{snewnew4}
\end{eqnarray} 
The jump of temperature is given by
\begin{eqnarray}
{1\over \eta_{cond}}-{1\over \eta_{gas}}\sim {\mu^{2/3}\over (\ln\mu)^{7/3}}. 
\label{wq3}
\end{eqnarray}  

Finally, we can study the behaviour of the fraction of mass contained in the germ and in the condensate  as $\mu\rightarrow +\infty$ for a given value of energy $\Lambda$. For the condensate (points C), $\alpha\rightarrow 0$ and $\eta\rightarrow 0$ so that Eqs. (\ref{c13}) (\ref{c14}) simplify in
\begin{eqnarray}
{3\over 2\eta}={3\over 7}\lambda\mu^{2/3}\alpha^{7/3},
\label{s6}
\end{eqnarray}  
\begin{eqnarray}
{3\over 2}\ln\eta+\lambda\mu^{2/3}\alpha^{4/3}\eta=\ln\mu.
\label{s7}
\end{eqnarray}
The first equality simply means that the potential energy of the core
tends to $-\infty$ so that the temperature of the halo must rise to
$+\infty$ so as to maintain the total energy fixed. Combining
Eqs. (\ref{s6})(\ref{s7}) and using Eq. (\ref{c15}), we obtain
\begin{eqnarray}
\alpha_{cond}\sim {1\over\ln\mu},\qquad  \eta_{cond}\sim {(\ln\mu)^{7/3}\over \mu^{2/3}}, \qquad S_{cond}\sim \ln\mu.
\label{s8}
\end{eqnarray}
Since the entropy diverges as $\mu\rightarrow +\infty$, we recover the well-known fact that there is no global entropy maximum for a classical self-gravitating gas. Note that the divergence of entropy is relatively slow (logarithmic). For the germ (points B), we have in the classical limit
\begin{eqnarray}
\alpha_{germ}\rightarrow 0,\qquad \eta_{germ}\rightarrow\eta_{gas},\qquad S_{germ}\rightarrow S_{gas}.
\label{s9}
\end{eqnarray}
These results simply reflect the fact that the unstable branch approaches the gaseous branch as $\mu\rightarrow +\infty$ but still differs from it by a presence of a small germ. The size of the germ is determined by Eq. (\ref{c14}). For $\mu\rightarrow +\infty$, it leads to 
\begin{eqnarray}
\lambda\mu^{2/3}\alpha^{4/3}\eta_{gas}=\ln\mu
\label{snew5}
\end{eqnarray}
so that
\begin{eqnarray}
\alpha_{germ}\sim {(\ln\mu)^{3/4}\over\mu^{1/2}}.
\label{s10}
\end{eqnarray}
For a given energy, the mass contained in the condensate and in the germ goes to zero as we approach the classical limit. However, the relation
\begin{eqnarray}
{\alpha_{germ}\over \alpha_{cond}}\sim {(\ln\mu)^{7/4}\over\mu^{1/2}}\ll 1,
\label{s11}
\end{eqnarray}
indicates that the size of the germ is much smaller than the size of the condensate. 

These results are consistent with the proof given in Ref. \cite{crs} for the absence of global entropy maximum in the microcanonical ensemble. In words, we can make the entropy diverge by approaching an arbitrarily small fraction of particles in the core ($M_{*}\ll M$) so that the potential energy goes to $-\infty$. Since the total energy is conserved, the temperature must rise to $+\infty$ and this leads to a logarithmic divergence of the entropy. This ``natural'' evolution is confirmed by dynamical models of self-gravitating systems (see the discussion in Ref. \cite{crs}). It is found that the gravitational collapse of classical point masses leads to a finite time singularity (the central density becomes infinite in a finite time $t_{coll}$) with a slow algebraic divergence of the temperature and a logarithmic divergence of the entropy as $t\rightarrow t_{coll}$. In addition, the mass contained in the core tends to zero as $t\rightarrow t_{coll}$ (the density profile at $t=t_{coll}$ is close to a power law $\rho\simeq r^{-\alpha}$, with $\alpha\simeq 2.2<3$) in agreement with our previous observations.

We now consider the canonical situation. In the limit $\mu\rightarrow +\infty$, the phase transition occurs close to maximum temperature of the condensed phase, corresponding to $\eta_{*}(\mu)$.  Using Eq. (\ref{c14}), we find that the condition of temperature maximum $d\eta=0$ is equivalent to
\begin{eqnarray}
{4\over 3}\lambda\mu^{2/3}\alpha^{1/3}\eta-{1\over 1-\alpha}-{9\over 5}\eta=0.
\label{s12}
\end{eqnarray} 
This equation, together with Eq. (\ref{c14}), determines the function $\eta_{*}(\mu)$. In the limit $\mu\rightarrow +\infty$, the fraction of mass contained in the condensate comes close to unity and the maximum temperature increases. Taking the limit  $\alpha\rightarrow 1$ and $\eta\rightarrow 0$ in Eq. (\ref{s12}), we get 
\begin{eqnarray}
{4\over 3}\lambda\mu^{2/3}\eta={1\over 1-\alpha}.
\label{s13}
\end{eqnarray}  
Simplifying Eqs. (\ref{c13})(\ref{c14}) in the same approximation,  we obtain
\begin{eqnarray}
\Lambda={3\over 7}\lambda\mu^{2/3}-{3\over 2}(1-\alpha){1\over \eta},
\label{s14}
\end{eqnarray}  
\begin{eqnarray}
\ln(1-\alpha)+{3\over 2}\ln\eta+\lambda\mu^{2/3}\eta=\ln\mu.
\label{s15}
\end{eqnarray}  
We find therefore that the parameters at the point of maximum temperature behave like
\begin{eqnarray}
1-\alpha_{*}\sim {1\over\ln\mu},\qquad \eta_{*}\sim {\ln\mu\over \mu^{2/3}},\qquad \Lambda_{*}-\Lambda_{max}(\mu)\sim -{\mu^{2/3}\over (\ln\mu)^{2}}.
\label{s16}
\end{eqnarray}

The transition point $\eta_{t}(\mu)$ in the canonical ensemble is obtained
by equating the free energy of the two phases. Using the results of Sec. \ref{sec_gas} and Sec. \ref{sec_cp}, we obtain the general relation
\begin{eqnarray}
-{3\over 2}\ln\eta+\ln\mu+C+1+{3\over 5}\eta=(1-\alpha)\biggl\lbrack -{3\over 2}\ln\eta-\ln(1-\alpha)+\ln\mu+C+{5\over 2}\biggr\rbrack\nonumber\\
+ {3\over 7}\lambda\eta\mu^{2/3}\alpha^{7/3}-{3\over 2}(1-\alpha)+{3\over 2}\eta\alpha(1-\alpha)+{3\over 5}\eta(1-\alpha)^{2}, 
\label{s17}
\end{eqnarray}   
where $\alpha$ is the mass contained in the condensate at the
transition point.  In the limit $\mu\rightarrow +\infty$ and
$\alpha\rightarrow 1$, this relation reduces to
\begin{eqnarray}
-{3\over 2}\ln\eta+\ln\mu= {3\over 7}\eta\lambda\mu^{2/3}. 
\label{s18}
\end{eqnarray}   
The free energy of the gaseous phase (l.h.s.) is dominated by the contribution of the entropy and the free energy of the condensed phase (r.h.s.) is dominated by the energy of the core. We find therefore that the temperature of transition behaves like
\begin{eqnarray}
\eta_{t}\sim {\ln\mu\over \mu^{2/3}}.
\label{s19}
\end{eqnarray}  
The mass contained in the condensate is determined by Eq. (\ref{c14}). In the limit $\mu\rightarrow +\infty$, it simplifies in 
\begin{eqnarray}
\ln(1-\alpha)+{3\over 2}\ln\eta+\lambda\mu^{2/3}\eta=\ln\mu.
\label{sneww1}
\end{eqnarray}  
Solving for $\alpha$ in Eqs. (\ref{s18})(\ref{sneww1}), we get
\begin{eqnarray}
1-\alpha\sim {1\over \mu^{8/3}}.
\label{s20}
\end{eqnarray}  
According to Eqs. (\ref{g4})(\ref{c13}) (\ref{s19}), the energy of the
gaseous phase and the energy of the condensed phase behave at the
transition point like
\begin{eqnarray}
\Lambda_{gas}\sim -{\mu^{2/3}\over \ln\mu},\qquad \Lambda_{cond}-\Lambda_{max}(\mu)\sim -{1\over \mu^{2}\ln\mu}.
\label{sneww2}
\end{eqnarray} 
The jump of energy (latent heat) is 
\begin{eqnarray}
\Lambda_{cond}-\Lambda_{gas}\sim \mu^{2/3}\biggl (1+{3\over 4 \ln\mu}\biggr ).
\label{s21}
\end{eqnarray}  

Finally, let us determine the fraction of mass contained in the condensate and in the germ as $\mu\rightarrow +\infty$ for a given value of temperature.  For the condensate (points C), $\alpha\rightarrow 1$ so that  Eq. (\ref{c14}) leads to 
\begin{eqnarray}
1-\alpha_{cond}\sim e^{-\lambda\eta\mu^{2/3}}.
\label{s22}
\end{eqnarray} 
The free energy of the condensed state behaves like  
\begin{eqnarray}
J\sim \Lambda_{max}(\mu)\eta\sim \mu^{2/3}.
\label{s23}
\end{eqnarray} 
The divergence of the free energy in the canonical ensemble is more rapid than the divergence of the entropy in the microcanonical ensemble. This is simply because the free energy is dominated by the divergence of the (potential) energy while the entropy is dominated by the divergence of the logarithm of the temperature. For the germ (points B), we have in the classical limit 
\begin{eqnarray}
\alpha_{germ}\rightarrow 0\qquad \Lambda\rightarrow \Lambda_{gas}\qquad J\rightarrow J_{gas}.
\label{s24}
\end{eqnarray} 
Using Eq. (\ref{c14}), we find that
\begin{eqnarray}
\alpha_{germ}\sim {(\ln\mu)^{3/4}\over\mu^{1/2}}.
\label{s25}
\end{eqnarray}
In the limit $\mu\rightarrow +\infty$, the germ contains almost no mass while the condensate contains almost all the mass. 

These results again agree with the proof given in Ref. \cite{chaviso} for the absence of a global maximum of free energy in the canonical ensemble. In words, we can make the free energy diverge by collapsing the mass $M$ to a point (the divergence of $J$ is simply due to the divergence of the potential energy; the entropy has a weak, logarithmic, negative divergence). This argument would suggest that the natural evolution of a system of classical point masses in the canonical ensemble is to develop a density profile in the form of a $\delta$-function with all the mass at $r=0$. This is {\it not} what numerical simulations of gravitational collapse show (see the discussion in Ref. \cite{crs}). It is usually found that when the system is held at a fixed temperature, the self-similar collapse leads to a density profile close to the power law $\rho\sim r^{-2}$ at $t=t_{coll}$. This profile has a vanishing mass at $r=0$ and its free energy is not divergent. Therefore, a finite time singularity prevents the system from reaching arbitrarily large values of the free energy . It is not known whether other solutions of these dynamical equations (not necessarily self-similar) can lead to the expected $\delta$-function with $J=+\infty$.

\section{Classical gas with a short distance cut-off}
\label{sec_cg}

The case of a self-gravitating gas with a short distance cut-off  was first considered by Aronson \& Hansen \cite{aronson} and more recently by Stahl {\it et al.} \cite{stahl} (see discussion in Sec. \ref{sec_comp}). The equilibrium phase diagram of this system is similar to the one obtained for self-gravitating fermions (see Figs. \ref{le5}-\ref{fel}). Indeed, the degeneracy parameter $\mu$ plays the same role as the inverse of the short distance cut-off $a$. The interpretation of the phase transitions is therefore similar but the dependance of the critical parameters with the cut-off is different. We shall therefore reformulate our analytical model to the case of a classical hard spheres gas and determine how the previous results are modified in this new situation.

\subsection{The ``gaseous'' phase}
\label{sec_hg}

We model the gaseous phase by a uniform distribution of matter occupying the whole container. The energy and the entropy are therefore given by
\begin{eqnarray}
E={3\over 2}NT-{3GM^{2}\over 5R},
\label{hg1}
\end{eqnarray}  
\begin{eqnarray}
S/N={3\over 2}+{3\over 2}\ln \biggl ({2\pi T\over m}\biggr )-\ln\biggl ({3N\over 4\pi R^{3}}\biggr ).
\label{hg2}
\end{eqnarray} 
In dimensionless variables, these equations can be rewritten
\begin{eqnarray}
\Lambda={3\over 5}-{3\over 2\eta},
\label{hg3}
\end{eqnarray}  
\begin{eqnarray}
S/N=-{3\over 2}\ln\eta.
\label{hg4}
\end{eqnarray} 
In the expression (\ref{hg4}) for the entropy, we have not written the constant term
\begin{eqnarray}
{3\over 2}+{3\over 2}\ln (2\pi)+\ln\biggl\lbrack {4\pi R^{3}\over 3N}\biggl ({GM\over R}\biggr )^{3/2}\biggr\rbrack,
\label{hg5}
\end{eqnarray} 
which plays no role in the problem.

\subsection{The ``condensed'' phase}
\label{sec_hc}

We model the ``condensed'' phase by a nucleus and an atmosphere, each of uniform density. The velocity distribution of the particles is assumed to be Maxwellian with temperature $T$.  Let $R_{*}$ be the radius of the nucleus and $N_{*}$ the number of particles that it contains. We introduce an excluded volume $\sim {4\over 3}\pi a^{3}$ around each particle, where $a$ can be regarded as the ``effective'' size of the particles. $R_{*}$ and $N_{*}$ are therefore related to each other by a relation of the form
\begin{eqnarray}
R_{*}^{3}=4 g N_{*}a^{3},
\label{hc1}
\end{eqnarray} 
where $g$ is a geometrical factor with order of magnitude unity which depends on the nature of the close-packing (see Ref. \cite{aronson}). The energy of the core is
\begin{eqnarray}
E_{*}={3\over 2}N_{*}T-{3GM_{*}^{2}\over 5R_{*}},
\label{hc2}
\end{eqnarray} 
and its entropy
\begin{eqnarray}
S_{*}=N_{*}\biggl\lbrack {3\over 2}+{3\over 2}\ln \biggl ({2\pi T\over m}\biggr )-\ln\biggl ({3N_{*}\over 4\pi R_{*}^{3}}\biggr )\biggr\rbrack.
\label{hc3}
\end{eqnarray} 
For the halo, we have
\begin{eqnarray}
E_{halo}={3\over 2}(N-N_{*})T-{3GM_{*}(M-M_{*})\over 2R}-{3G(M-M_{*})^{2}\over 5R},
\label{hc4}
\end{eqnarray} 
\begin{eqnarray}
S_{halo}=(N-N_{*})\biggl\lbrace {3\over 2}+{3\over 2}\ln \biggl ({2\pi T\over m}\biggr )-\ln\biggl \lbrack {3(N-N_{*})\over 4\pi R^{3}}\biggr \rbrack\biggr\rbrace.
\label{hc5}
\end{eqnarray} 
Like in the case of fermions, we have considered that the volume of the nucleus is much smaller than the volume of the halo. Adding these expressions and using Eq. (\ref{hc1}) to express the radius of the core as a function of its mass $M_{*}=N_{*}m$, we obtain for the whole configuration
\begin{eqnarray}
E={3\over 2}NT-{3Gm^{1/3}M_{*}^{5/3}\over 5a}-{3GM_{*}(M-M_{*})\over 2R}-{3G(M-M_{*})^{2}\over 5R},
\label{hc6}
\end{eqnarray} 
\begin{eqnarray}
\qquad S=N\biggl\lbrack {3\over 2}+{3\over 2}\ln \biggl ({2\pi T\over m}\biggr )-\ln\biggl ({3N\over 4\pi R^{3}}\biggr )\biggr\rbrack
-N_{*}\ln\biggl ({R^{3}\over Na^{3}}\biggr )-(N-N_{*})\ln\biggl ({N-N_{*}\over N}\biggr ).
\label{hc7}
\end{eqnarray} 
Let  $\alpha=M_{*}/M$ denote the fraction of mass contained in the nucleus. We also introduce the filling factor
\begin{eqnarray}
\mu={R\over a (4gN)^{1/3}},
\label{hc8}
\end{eqnarray}  
which can be regarded as an inverse normalized hard spheres radius. The case of point masses corresponds to the limit $\mu\rightarrow +\infty$. Clearly, $\mu$  plays the same role as the degeneracy parameter in section \ref{sec_para}. In dimensionless form, the equations of the problem become
\begin{eqnarray}
\Lambda=-{3\over 2\eta}+{3\over 5}\mu\alpha^{5/3}+{3\over 2}\alpha (1-\alpha)+{3\over 5}(1-\alpha)^{2},
\label{hc9}
\end{eqnarray}  
\begin{eqnarray}
S/N=-{3\over 2}\ln\eta-3\alpha\ln\mu-(1-\alpha)\ln (1-\alpha),
\label{hc10}
\end{eqnarray}
where we have again eliminated the constant (\ref{hg5}) from the
expression of the entropy.  We now determine the mass of the nucleus
by maximizing the entropy (\ref{hc10}) at fixed energy. This yields
the relation
\begin{eqnarray}
1+\ln(1-\alpha)-3\ln\mu=-{3\eta\over 10}+{9\over 5}\alpha\eta-\mu\eta\alpha^{2/3}.
\label{hc11}
\end{eqnarray}
Eqs. (\ref{hc9})(\ref{hc10})(\ref{hc11}) determine the equilibrium phase diagram of a classical hard spheres gas in the framework of our analytical model. The description of this diagram (and its dependance on the parameter $\mu$) is similar to the one given in section \ref{sec_analytical} for self-gravitating fermions. The minimum energy corresponds to the configuration for which all the mass is in the nucleus at zero temperature. Taking the limit $\alpha\rightarrow 1$ and $\eta\rightarrow +\infty$ in Eq. (\ref{hc9}), we get
\begin{eqnarray}
\Lambda_{max}(\mu)={3\over 5}\mu.
\label{hc12}
\end{eqnarray} 
For $\Lambda\rightarrow \Lambda_{max}$ and for sufficiently large values of $\mu$, Eqs. (\ref{hc9})(\ref{hc11}) yield
\begin{eqnarray}
1-\alpha\sim e^{-\mu\eta}, \qquad {\Lambda_{max}-\Lambda}\sim {3\over 2\eta}
\label{hc13}.
\end{eqnarray} 
Note that the relation between the temperature and the energy is different from the corresponding one for self-gravitating fermions (see Eq. (\ref{c18})).

\subsection{Scaling laws in the limit $\mu\rightarrow +\infty$}
\label{sec_hslaws}

The derivation of the scaling laws for the critical parameters of a
classical hard spheres gas is essentially the same as for
self-gravitating fermions (section \ref{sec_scaling}). We shall
directly give the results without detailed discussion.

In the microcanonical ensemble, the point of maximum
energy, corresponding to $\Lambda_{*}(\mu)$, is determined by the
relation
\begin{eqnarray}
{2\over 3}\eta^{2}\biggl ({3\over 10}-{9\over 5}\alpha+\mu\alpha^{2/3}\biggr )^{2}={2\over 3}\mu\eta\alpha^{-1/3}-{9\over 5}\eta-{1\over 1-\alpha}.
\label{w1}
\end{eqnarray} 
In the limit $\mu\rightarrow +\infty$, the thermodyanical parameters behave at the point of maximum energy like
\begin{eqnarray}
\alpha_{*}\sim {1\over\ln\mu},\quad \eta_{*}\sim {(\ln\mu)^{5/3}\over \mu},\quad \Lambda_{*}\sim -{\mu\over (\ln\mu)^{5/3}}.  
\label{w2}
\end{eqnarray}

At the transition point $\Lambda_{t}(\mu)$, the equality of the entropy of the two phases leads to the relation
\begin{eqnarray}
-{3\over 2}\ln\eta_{cond}-3\alpha\ln\mu-(1-\alpha)\ln(1-\alpha)=-{3\over 2}\ln\eta_{gas}.
\label{w3}
\end{eqnarray} 
In the limit $\mu\rightarrow +\infty$, the fraction of mass contained in the condensate behaves like
\begin{eqnarray}
\alpha\sim {1\over \ln \mu}.
\label{w4}
\end{eqnarray} 
The energy of transition and the temperature of each phase are given by
\begin{eqnarray}
\Lambda_{t}\sim  -{\mu\over (\ln\mu)^{5/3}},\qquad \eta_{gas}\sim \eta_{cond}\sim {(\ln\mu)^{5/3}\over \mu}.  
\label{w5}
\end{eqnarray} 
The jump of temperature at the transition point is
\begin{eqnarray}
{1\over \eta_{cond}}-{1\over \eta_{gas}}\sim {\mu\over (\ln\mu)^{5/3}}. 
\label{w5bis}
\end{eqnarray}

For a given energy,  the thermodynamical parameters of the condensate behave, in the limit $\mu\rightarrow +\infty$, like
\begin{eqnarray}
\alpha_{cond}\sim {1\over\ln\mu},\qquad \eta_{cond}\sim {(\ln\mu)^{5/3}\over \mu}, \qquad S_{cond}\sim\ln\mu. 
\label{w6}
\end{eqnarray}
For the germ, we have 
\begin{eqnarray}
\alpha_{germ}\sim \biggl ({\ln\mu\over\mu}\biggr )^{3/2}, \qquad \eta_{germ}\rightarrow \eta_{gas},\qquad S_{germ}\rightarrow S_{gas}. 
\label{w7}
\end{eqnarray}
 
In the canonical ensemble, the maximum temperature of the condensed phase, corresponding to $\eta_{*}(\mu)$, is determined by the relation
\begin{eqnarray}
{2\over 3}\mu\eta\alpha^{-1/3}-{1\over 1-\alpha}-{9\over 5}\eta=0. 
\label{w8}
\end{eqnarray}
In the limit $\mu\rightarrow +\infty$, the thermodyanical parameters behave at the point of maximum temperature like 
\begin{eqnarray}
1-\alpha_{*}\sim {1\over\ln\mu},\quad \eta_{*}\sim {\ln\mu\over \mu},\quad \Lambda_{*}-\Lambda_{max}(\mu)\sim -{\mu\over \ln\mu}.  
\label{w9}
\end{eqnarray}

At the transition point  $\eta_{t}(\mu)$, the equality of the free energy of the two phases yields
\begin{eqnarray}
{3\over 5}\eta=-3\alpha\ln\mu-(1-\alpha)\ln(1-\alpha)+{3\over 5}\mu\eta\alpha^{5/3}+{3\over 2}\eta\alpha (1-\alpha)+{3\over 5}\eta (1-\alpha)^{2}.
\label{w10}
\end{eqnarray}
In the limit $\mu\rightarrow +\infty$, the temperature of transition
behaves like
\begin{eqnarray}
\eta_{t}\sim {\ln\mu\over\mu},
\label{w11bis}
\end{eqnarray}
and the fraction of mass contained in the nucleus like
\begin{eqnarray}
1-\alpha\sim {1\over \mu^{2}}.
\label{w11}
\end{eqnarray}
The energy of the two phases at the transition point is given by
\begin{eqnarray}
\Lambda_{gas}\sim -{\mu\over\ln\mu},\qquad \Lambda_{cond}-\Lambda_{max}(\mu)\sim  -{\mu\over\ln\mu},
\label{w12}
\end{eqnarray}
and the jump of energy is
\begin{eqnarray}
\Lambda_{cond}-\Lambda_{gas}\sim\mu.
\label{w13}
\end{eqnarray}

For a given temperature, the thermodynamical parameters characterizing
the condensate behave, in the limit $\mu\rightarrow +\infty$, like
\begin{eqnarray}
1-\alpha_{cond}\sim e^{-\mu\eta}, \qquad \Lambda\sim\Lambda_{max}(\mu)\sim \mu, \qquad J\sim\eta\Lambda_{max}(\mu)\sim \mu.
\label{w14}
\end{eqnarray}
For the germ, we have
\begin{eqnarray}
\alpha_{germ}\sim \biggl ({\ln\mu\over\mu}\biggr )^{3/2},\qquad \Lambda\rightarrow \Lambda_{gas},\qquad J\rightarrow J_{gas}.
\label{E}
\end{eqnarray}

\subsection{Comparision with previous works}
\label{sec_comp}

Phase transitions in self-gravitating systems were first
investigated with the aid of toy models which could be solved exactly
without recourse to a meanfield approximation. For example,
Lynden-Bell \& Lynden-Bell \cite{ll} considered a system of $N$
particles confined to the surface of a sphere of variable radius. They
calculated exactly the density of states in the microcanonical ensemble
and showed the existence of a region with negative specific
heats. Then, they evaluated the partition function in the canonical
ensemble and demonstrated that the region of negative specific heats
is replaced by a remarkable giant phase transition connecting a
``gaseous'' phase (at high energies) to a ``condensed'' phase (at low
energies). Padmanabhan \cite{pad} obtained similar results with a
simpler model consisting on only two particles in gravitational
interaction confined within a spherical box. The phase diagram
determined by these authors is similar to the one reported in
Fig. \ref{fel}. These models exhibit a phase transition in the
canonical ensemble but not in the microcanonical ensemble (unlike in
Fig. \ref{le5}). This is because, for these simple models, the
density of states remains finite when the small-scale cut-off $a$ is
set equal to zero whereas for more realistic self-gravitating systems,
it diverges. By contrast, the partition function is divergent for
$a=0$ and this leads to the occurence of a phase transition in the
canonical ensemble when $a$ is sufficiently small. Padmanabhan
investigated the dependance of the critical parameters with the small
scale cut-off $a$. In particular, he found that the temperature of
transition is given by
\begin{eqnarray}
T_{t}={Gm^{2}\over 3 a \ln(R/a)}.
\label{comp1}
\end{eqnarray}
This expression qualitatively agrees with our result (\ref{w11bis}), which becomes in dimensional variables
\begin{eqnarray}
T_{t}\sim{GN^{2/3}m^{2}\over  a \ln(R/aN^{1/3})}.
\label{comp2}
\end{eqnarray}
Recall that $N=2$ in Padmanabhan's model. He also computed the change of energy at the transition point and found that
\begin{eqnarray}
E_{gas}-E_{cond}={Gm^{2}\over a}\biggl (1-{1\over 3\ln (R/a)}\biggr ).
\label{comp3}
\end{eqnarray}
This expression also qualitatively agrees with our result (\ref{w13}),
leading to 
\begin{eqnarray}
E_{gas}-E_{cond}\sim {GN^{5/3}m^{2}\over a}.
\label{comp4}
\end{eqnarray}
The logarithmic correction in Eq. (\ref{comp3}) is a particularity of
Padmanabhan's model arising from the low value of $N$. Finally,
Padmanabhan investigated the dependance of the $T(E)$ curve with the
small-scale cut-off $a$ (see his Fig. 3.2). For large $a$, his diagram
is similar to that of Fig. \ref{multimu}. In particular, there exists
a critical short distance cut-off above which the phase transition
disappears (tricritical point). It should be stressed that the
statistical approach based on the evaluation of $g(E)$ or $Z(\beta)$
does not determine the metastable states unlike the thermodynamical
approach based on the maximization of $S\lbrack \rho\rbrack$ or
$J\lbrack\rho\rbrack$. Only the true equilibrium states, which
correspond to global maxima of $S$ or $J$, appear in the $T(E)$
diagram (in other words, we directly obtain the ``plateaux'' without
beeing required to make a Maxwell construction). These equilibrium
states are expected to be reached for $t\rightarrow +\infty$ but they
are not necessarily the most relevant for astrophysical applications:
as discussed previously, the metastable equilibrium states may be
long-lived and may correspond to the structures that are actually
observed in the universe. Indeed, the statistical mechanical approach
tells nothing about the timescales involved in the establishement of
the equilibrium.

\begin{figure}[htbp]
\centerline{
\psfig{figure=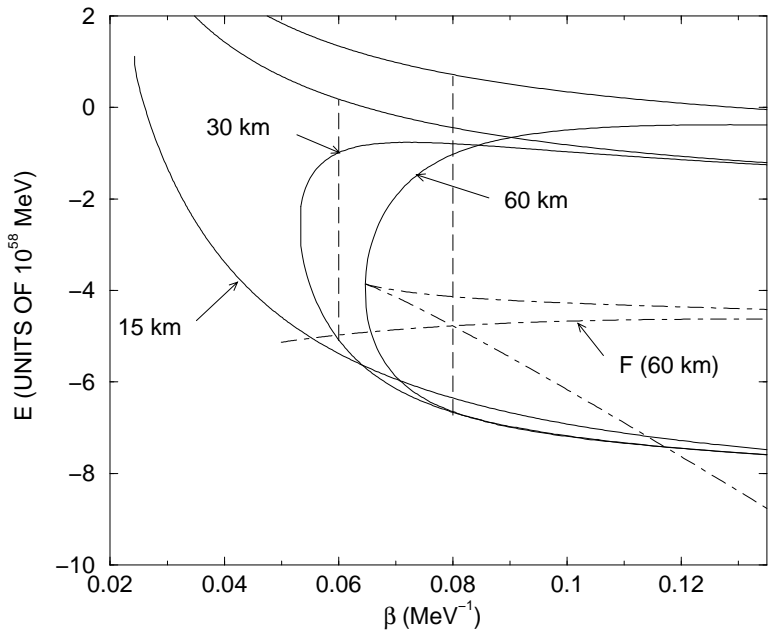,angle=0,height=8.5cm}}
\caption{Equilibrium phase diagram for a hard spheres gas obtained in the framework of our analytical model. The parameters have been chosen so as to correspond to Fig. 2 of Aronson \& Hansen \cite{aronson}: $R=60\ {\rm km}$, $R=30\ {\rm km}$, $R=15\ {\rm km}$; $N=10^{57}$; $m={\rm neutron \ mass}=1.67\times 10^{-24}\ {\rm g}$; $a=0.4\times 10^{-13}\ {\rm cm}$. These parameters correspond to a neutron-star-like structure. In our analytical model, we have adopted a value of the geometrical factor $g=2$ \cite{aronson}. The agreement with the numerical study of Aronson \& Hansen is fairly good. For small radius $R=15\ {\rm km}$, there is no phase transition. For larger radii ($R=30\ {\rm km}$, $R=60\ {\rm km}$), a phase transition connects the ``gaseous'' phase (upper branch) to the ``condensed'' phase (lower branch).  This phase transition forms a Maxwell plateau (dashed-line) at a temperature depending on the size $R$ of the system (according to Eqs. (\ref{w10}) and (\ref{hc11})). In dot-dashed line, we have ploted the free energy $F=E-TS$ as a function of the inverse temperature for the case $R=60\ {\rm km}$. The crossing point determines the temperature of transition.}
\label{hansen}
\end{figure}

Phase transitions in self-gravitating systems have also been
investigated in the meanfield approximation for less idealized
models. Aronson \& Hansen \cite{aronson} have considered the case of a
classical hard spheres gas modeled by a van der Waals equation of
state. They considered a relatively large cut-off and evidenced only
the phase transition in the canonical ensemble (they obtained a
diagram similar to that of Fig. \ref{fel}). They also proposed a
simple analytical model to describe this phase transition. In their
model, the ``gaseous'' phase consists of $N$ particles spread with
uniform density throughout the whole container and the ``condensed''
phase has {\it all} $N$ particles collapsed into a central core of
uniform density. The model that we proposed in Sec. \ref{sec_cg} is
more general because we allow the condensate to contain an arbitrary
fraction of the total mass. Then, the fraction of mass that is actually
achieved at equilibrium is {\it determined} by maximizing the free
energy versus $\alpha$. In the canonical ensemble (the only situation
discussed by Aronson \& Hansen), the fraction of mass contained in the
condensate is close to one, so that our precedure provides additional
support to their {\it Ansatz}. However, our model allows us to
describe also the unstable solution with the ``germ'' ($\alpha\ll 1$)
and to obtain a better representation of the whole bifurcation diagram
(see Fig. \ref{hansen}). In addition, our model can also describe the
phase transitions in the microcanonical ensemble (occuring for
sufficiently small values of the cut-off $a$) which was not considered
by Aronson and Hansen.

The study of Aronson \& Hansen  was reconsidered by Stahl {\it et al.} \cite{stahl} who pointed out that the van der Waals equation of state does not adequately describe hard spheres systems at high densities. They considered a more general equation of state and studied in detail the case of small filling factors for which a phase transition occurs in the microcanonical ensemble. They also determined numerically the dependance of the transition temperature (in the canonical ensemble) as a function of the filling factor and considered applications of their results in the context of planet formation.

Following the works of Aronson and Hansen \cite{aronson} and Stahl
{\it et al.} \cite{stahl}, different authors have attempted to
describe phase transitions in self-gravitating systems by introducing
a small scale regularization of the gravitational potential. For
example, in the study of Follana \& Laliena \cite{laliena}, the
softening is achieved by truncating to $N$ terms an expansion of the
Newtonian potential in spherical Bessel functions. These authors
obtained an equilibrium phase diagram similar to the one in
Fig. \ref{fel}. However, they could not achieve very large values of
$N$ in their study so that they were unable to see the development of
the spiral (and the corresponding phase transition in the
microcanonical ensemble) as $N\rightarrow +\infty$. Another analysis
of phase transitions in self-gravitating systems was provided by
Youngkins \& Miller \cite{miller} with a one-dimensional model of
concentric spherical mass shells. They studied this system in the
microcanonical, canonical and grand canonical ensembles both
numerically and analytically in the meanfield approximation. They
found an overall good agreement between their numerical simulations
and the meanfield predictions. They observed phase transitions in the
microcanonical and canonical ensembles but not in the grand canonical
ensemble in which the system remains homogeneous. This last result may
be, however, an artefact of their one-dimensional model. It is
plausible that, in the grand canonical ensemble, the self-gravitating
gas fragments in a series of clumps (at different scales) as observed
in cosmology and for the interstellar medium. Some theoretical
arguments in favour of this scenario have been given by Semelin {\it
et al.} \cite{semelin2} and Chavanis \cite{chaviso}. Of course, to
study the development of these clumps, it is necessary to extend the
thermodynamical analysis to the full three dimensional problem and
relax the assumption of spherical symmetry.

The thermodynamics of self-gravitating fermions was investigated by
Lynden-Bell \& Wood and Hertel \& Thirring in the early seventies (see
the discussion of Aronson \& Hansen \cite{aronson}), but these papers were
apparently not published. In Refs. \cite{hertel,robert}, it is proved
that a rigorous thermodynamic limit exists for self-gravitating
fermions but the corresponding phase diagram is not explicitly
given. This equilibrium phase diagram was calculated by Bilic \& Viollier
\cite{bilic} for a particular value of the degeneracy parameter
adapted to a cosmological settling. It was also calculated
independantly by Chavanis \& Sommeria \cite{cs} for an arbitrary
degree of degeneracy in the context of the theory of ``violent
relaxation''. In Ref. \cite{cs}, the development of the spiral for
high values of the degeneracy parameter and the associated phase
transition that occurs in the microcanonical ensemble (in addition to
the more well-known phase transition in the canonical ensemble) was
clearly shown. However, the point of transition was not explicitly
determined and this has been done in the present study.

\subsection{Astrophysical applications}
\label{sec_astro}

The application of the hard spheres model in astrophysics could
concern the fragmentation of the interstellar medium and the formation
of stars or even smaller objects such as planets \cite{stahl}. These
objects would correspond to the ``condensate'' that results from the
collapse of a cloud of gas or dust. On the other hand, the model of
self-gravitating fermions could have applications for massive
neutrinos in dark matter models \cite{ruffini,bilic}, white dwarfs
\cite{chandra} and neutron stars \cite{oppenheimer,hertel}. By cooling
below a critical temperature, a condensed phase emerges consisting of
a completely degenerate nucleus surrounded by a dilute enveloppe, as
extensively studied in early models of stellar structure
\cite{chandra}. This model could also be relevant for the ``violent
relaxation'' of collisionless stellar systems \footnote{In fact, the
problem is complicated because violent relaxation eventually fades
before the maximum entropy state is attained. Thus, equation
(\ref{fd6}) [or Eq. (\ref{fd7})] is unlikely to be reached throughout
the whole cluster. However, it is reasonable to hold in the central
region in which violent relaxation occurs most violently
\cite{lb,hjorth}.} such as elliptical galaxies
\cite{lb,cs,dubrovnik}. In that case, the exclusion principle
is a consequence of the Liouville theorem. Since degeneracy can
stabilize the system without changing its overall structure at large
distances, we have suggested in Ref. \cite{cs} that degeneracy could
play a role in galactic nuclei. The recent simulations of Leeuwin \&
Athanassoula
\cite{leeuwin} and the theoretical model of Stiavelli \cite{stiavelli}
are consistent with this idea especially if the nucleus of elliptical
galaxies contains a primordial massive black hole. Indeed, the effect
of degeneracy (in the sense of Lynden-Bell) on the distribution of
stars surrounding the black hole can explain the cusps observed at the
center of galaxies. Whether or not elliptical galaxies are degenerate
remains however a matter of debate because when the core becomes
dense, two-body encounters will come into play and break the Liouville
theorem (Shu's criticism \cite{shu}). This form of degeneracy may,
however, be relevant for massive neutrinos in Dark Matter models where
it competes with quantum degeneracy \cite{kull}. In fact, the thermal
equilibrium distribution of massive neutrinos in Dark Matter models
might be justified more by the process of ``violent relaxation'' than
by a collisional relaxation. Indeed, the time scale of gravitational
two-body encounters for neutrinos is extremely long so that the
criticism raised by Shu does not apply \cite{madsen,kull}. Therefore,
the commonly adopted Fermi-Dirac distribution of self-gravitating
neutrinos might be due to Lynden-Bell's type of degeneracy rather than
to quantum mechanics. Anyway, whatever the source of the exclusion
principle (Lynden-Bell or Pauli), the self-gravitating Fermi-Dirac
model predicts the formation of a dense degenerate nucleus (``fermion
star'') with a small radius and a large mass
\cite{bilic,cs}. As suggested in Refs. \cite{bilic,cs,bilic2}, this
dense degenerate nucleus could be an alternative to black holes at the
center of galaxies. On the other hand, at large distances, the density
of the (isothermal) self-gravitating Fermi gas decays like $r^{-2}$
which is a condition that dark galactic halos must fulfill in order to
reproduce the flat rotation curves of spiral galaxies
\cite{bt}. Therefore, this model of self-gravitating fermions 
has a chance to account for the structure of dark matter in galactic
halos.

\section{Conclusion}
\label{sec_conclusion}

In this paper, we have described the inequivalence of statistical
ensembles and the nature of phase transitions in self-gravitating
systems by considering the case of self-gravitating fermions or the
case of a classical hard spheres gas. The introduction of an effective
repulsion at short distances avoids the singularity of the ``naked''
gravitational potential. It is likely that similar results will be
obtained with different forms of the regularization. For large values
of the cut-off $a$, there are no phase transition. For intermediate
values of $a$, phase transtions occur in the canonical ensemble but
not in the microcanonical ensemble. The corresponding phase diagram is
of the type of Fig. \ref{fel} and has been found by various authors
\cite{aronson,bilic,cs,laliena}. For smaller values of $a$, phase
transitions occur both in the canonical and in the microcanonical
ensemble. The corresponding phase diagram is of the type of
Fig. \ref{le5} and was first obtained in Ref. \cite{cs}. As the
small-scale cut-off is decreased, the $T(E)$ curve winds up and tends
to the classical spiral for $a\rightarrow 0$ (see Fig. \ref{etalambda}).

Depending on the value of the cut-off $a$ and of the ensemble
considered (microcanonical or canonical), three kinds of phase
transitions can be evidenced which separate a ``gaseous'' phase from a
``condensed'' phase. In the microcanonical ensemble and for
sufficiently small $a$ (Fig. \ref{le5}), a gravitational first order
phase transition  occurs at an energy $E_{t}(a)$ at which
the ``gaseous'' states pass from global to local entropy maxima and
the ``condensed'' states from local to global entropy maxima. This
transition is marked by a discontinuity of the temperature and of the
specific heats. In the limit $a\rightarrow 0$, the transition energy
is rejected to $+\infty$, so that the stable (gaseous) solutions of
the classical spiral are only {\it metastable}. These states may be
physical depending whether the initial condition lies in their
``basin of attraction'' or not \cite{cs,miller,crs}. However, the
metastable branch disappears at a critical energy $E_{c}$, discovered
by Antonov
\cite{antonov}, at which the ``gravothermal catastrophe'' \cite{lbw}
occurs. This collapse can be considered as a zeroth order phase
transition  since it is associated with a discontinuity of
entropy. For larger values of $a$, there is only one entropy maximum
for each value of energy (Fig \ref{fel}) and the previously described
phase transitions are suppressed. However, as we progressively decrease
energy, the self-gravitating gas acheives higher and higher density
contrasts and builds up a compact core containing more and more
mass. This gravitational ``clustering'' can be called a second order
phase transition  since the specific heats diverges at the
critical point  and the order parameter experiences a
rapid variation in the region of negative specific heats (but remains
continuous). Similar phase transitions occur in the
canonical ensemble. It is interesting that these phase transitions can
be understood with the aid of a simple analytical model which allows
one to determine the dependance of the thermodynamical parameters with
the cut-off value. The present study can be extended to include
rotation (in preparation). It is also important to develop non
equilibrium models to determine the structure of the ``basin of
attraction'' of the equilibrium states when several solutions exist. A
first step in that direction was made by Youngkins \& Miller
\cite{miller} and by Chavanis {\it et al.} \cite{crs} with the aid of
simplified dynamical models. These models could be used in particular
to investigate numerically the occurence of the phase transitions and
the robustness of the metastable states. If the system is placed in a
metastable state (local entropy maximum), 
it will eventually jump to the global entropy maximum, but this can
take an infinite (physically irrelevant) time. Indeed, the probability
that a fluctuation will allow the phase transition to develop is
expected to be extremely low (except maybe near the critical point)
\cite{newkatz}. Therefore, if the system is trapped in a metastable state 
(but still slowly evolving along the series of equilibrium by losing
mass or energy like for globular clusters), the phase transition will
occur at the critical point $\Lambda_{c}$ rather than at
$\Lambda_{t}$. However, if the system is initially far from
equilibrium, there is no simple criterion to decide a priori whether
it will converge towards the local or the global entropy maximum. Only
direct numerical simulations can answer this question and sketch the structure 
of the basin of attraction for self-gravitating systems.

\section{Acknowledgments}
\label{sec_ack}

I acknowledge interesting discussions with N. Bilic, T. Dauxois,
O. Fliegans, V. Laliena, D. Lynden-Bell, C. Sire and
J. Sommeria. Special thanks are due to J. Katz for useful comments on
this manuscript during his visit in Toulouse. I have also benefited 
from precious discussions with S. Ruffo, J. Barr\'e and F. Bouchet during
my stay at the University of Firenze (october 2001).
 
\newpage
\appendix

\section{Entropy of the self-gravitating Fermi gas}
\label{sec_entropy}

In this Appendix, we give the main steps for deriving the expression (\ref{p9}) for the entropy. Substituting the Fermi-Dirac distribution
\begin{eqnarray}
f={\eta_{0}\over 1+ke^{\psi}e^{\beta{v^{2}\over 2}}},
\label{ann1}
\end{eqnarray}
in the entropy (\ref{fd5}), we obtain after some rearrangements
\begin{eqnarray}
\eta_{0}S=M\ln k+\int\rho\psi d^{3}{\bf r}+K\beta+\eta_{0}\int\ln \biggl (1+{1\over k}e^{-\psi}e^{-\beta{v^{2}\over 2}} \biggr )d^{3}{\bf r}d^{3}{\bf v}. 
\label{ann2}
\end{eqnarray}
The last integral can be integrating by parts yielding the value $2\beta K/3$. Therefore, 
\begin{eqnarray}
\eta_{0}S=M\ln k+\int\rho\psi d^{3}{\bf r}+{5\over 3}K\beta.
\label{ann3}
\end{eqnarray}
Using the definition of $\psi$, we get
\begin{eqnarray}
\eta_{0}S=M\ln k+2\beta W-M\beta\Phi_{0}+{5\over 3}K\beta.
\label{ann4}
\end{eqnarray}
Now, the central density is determined by the relation $\psi(\alpha)=\beta(\Phi(R)-\Phi_{0})$ with $\Phi(R)=-GM/R$. Hence
\begin{eqnarray}
{S\eta_{0}\over M}=\ln k+{2\beta\over M} W+{\beta GM\over R}+\psi(\alpha)+{5\beta K\over 3M}.
\label{ann5}
\end{eqnarray}
Using Eq. (3.12) of Chavanis \& Sommeria \cite{cs} to express the potential energy $W$ and the kinetic energy $K=E-W$ in terms of $E$, we find
\begin{eqnarray}
{S\eta_{0}\over M}=\ln k+\eta+\psi(\alpha)-{7\over 3}\Lambda\eta-{2\over 9}\eta{\alpha^{10}\over \mu^{4}}I_{3/2} (ke^{\psi(\alpha)} ).
\label{ann6}
\end{eqnarray}
Finally, using $\eta=\mu^{2}/\alpha^{4}$, resulting from Eqs. (\ref{p5}) and 
(\ref{p7}), we obtain Eq. (\ref{p9}).

\end{document}